\documentclass[preprint2]{aastex}

\shorttitle{Spitzer spectra of the 2Jy sample }
\shortauthors{Dicken et al.}

\begin{document}

\title{Spitzer Mid-IR Spectroscopy of Powerful 2Jy and 3CRR Radio Galaxies. II. AGN Power Indicators and Unification}

\author{D. Dicken\altaffilmark{1},
C. Tadhunter\altaffilmark{2}, 
R. Morganti\altaffilmark{3,4}, 
D. Axon\altaffilmark{5,6},
A. Robinson\altaffilmark{5}, 
M. Magagnoli \altaffilmark{5}, 
P. Kharb\altaffilmark{7},
C. Ramos Almeida\altaffilmark{8,9},
B. Mingo\altaffilmark{10},
M. Hardcastle\altaffilmark{11},
N. P. H. Nesvadba\altaffilmark{12},
V. Singh\altaffilmark{12},
M.B.N. Kouwenhoven\altaffilmark{13},
M. Rose\altaffilmark{15}, 
H. Spoon\altaffilmark{15},
K. J. Inskip\altaffilmark{16},
J. Holt\altaffilmark{17}
 }

\altaffiltext{1}{CEA-Saclay, F-91191 Gif-sur-Yvette, France; daniel.dicken@cea.fr}
\altaffiltext{2}{University of Sheffield, Hounsfield Road, Sheffield, S3 7RH, UK;} 
\altaffiltext{3}{ASTRON, P.O. Box 2,7990 AA Dwingeloo The Netherlands}
\altaffiltext{4}{Kapetyn Astronmical Institute, University of Groningen, Postbuss 800, 9700 AV Groningen, The Netherlands}
\altaffiltext{5}{Rochester Institute of Technology, 84 Lomb Memorial Drive, Rochester NY14623, USA}
\altaffiltext{6}{University of Sussex, Pevensey 2, University of Sussex, Falmer, Brighton, BN1 9QH, UK} 
\altaffiltext{7}{Indian Institute of Astrophysics, II Block, Koramangala, Bangalore 560034, India}
\altaffiltext{8}{Instituto de Astrofisica de Canarias (IAC), C/V ia Lactea, s/n, E-38205, La Laguna, Tenerife, Spain} 
\altaffiltext{9}{Departamento de Astrofisica, Universidad de La Laguna, E-38205, La Laguna, Tenerife, Spain}
\altaffiltext{10}{Department of Physics \& Astronomy, University of Leicester, University Road, Leicester, LE1 7RH, United Kingdom}
\altaffiltext{11}{School of Physics, Astronomy and Mathematics, University of Hertfordshire, College Lane, Hatfield AL10 9AB} 
\altaffiltext{12}{Institut dÕAstrophysique Spatiale, CNRS, Universit\'e Paris Sud, 91405 Orsay, France}
\altaffiltext{13}{Kavli Institute for Astronomy and Astrophysics, Peking University, Yi He Yuan Lu 5, Haidian Qu, Beijing 100871, P.R.~China}
\altaffiltext{14}{Harvard Smithsonian Center for Astrophysics, 60 Garden St., Cambridge, MA 02138, USA}
\altaffiltext{15}{224 Space Sciences Building, Cornell University, Ithaca, NY 14853} 
\altaffiltext{16}{Max Planck Institute for Astronomy, K\"onigstuhl 17, 69117 Heidelberg, Germany}
\altaffiltext{17}{Leiden Observatory, Leiden University, PO Box 9513, 2300 RA Leiden, the Netherlands}

\begin{abstract}
It remains uncertain which continuum and emission line diagnostics best indicate the bolometric powers of active galactic nuclei (AGN), especially given the attenuation caused by the circum-nuclear material, and the possible contamination by components related to star formation. Here we use mid-IR spectra along with the multi-wavelength data to investigate the merit of various diagnostics of AGN radiative power, including the mid-IR [NeIII]$\lambda$25.89$\mu$m and [OIV]$\lambda$25.89$\mu$m fine structure lines, the optical [OIII]$\lambda5007$ forbidden line, and mid-IR 24$\mu$m, 5GHz radio, and X-ray continuum emission, for complete samples of 46 2Jy radio galaxies (0.05$<z<$0.7) and 17 3CRR FRII radio galaxies ($z<$0.1). We find that the mid-IR [OIV] line is the most reliable indicator of AGN power for powerful radio-loud AGN. By assuming that the [OIV] is emitted isotropically, and comparing the [OIII] and 24$\mu$m luminosities of the broad- and narrow-line AGN in our samples at fixed [OIV] luminosity, we show that the [OIII] and 24$\mu$m emission are both mildly attenuated in the narrow-line compared to the broad-line objects by a factor $\approx$2. However, despite this attenuation, the [OIII] and 24$\mu$m luminosities are better AGN power indicators for our sample than either the 5 GHz radio or the X-ray continuum luminosities]. We also detect the mid-IR 9.7$\mu$m silicate feature in the spectra of many objects but not ubiquitously: at least 40\% of the sample show no clear evidence for these features. We conclude that, for the majority of powerful radio galaxies, the mid-IR lines are powered by AGN photoionisation.

\end{abstract}


\section{Introduction}
\label{sec:intro}

Orientation-based unified schemes propose that certain components of the emission from AGN are radiated anisotropically (e.g. \citealp{urry95}). It is widely believed that part of this orientation dependence is caused by obscuring dust structures in the shape of tori or warped disks around the central AGN and broad-line regions (e.g. {\citealp{antonucci84}}; \citealp{nenkova08}; \citealp{Lawrence10}). This implies that, although AGN may have a common central engine, they can look drastically different in terms of their emission line spectra and broad-band continuum emission, as the orientation changes with respect to the line of sight. 

While the concept of obscuration-induced anisotropy is useful for our understanding of the relationship
between type 1 and type 2 AGN, the obscuring dust and gas can hinder our ability to determine key intrinsic
properties of AGN such as their total radiative (bolometric) powers. In particular, the degree to which different
optical emission components suffer from obscuration is uncertain. Therefore, it is important to identify alternative bolometric AGN power indicators that do not suffer from attenuation. This is necessary both to further test unification ideas and to improve our understanding of the AGN population as a whole (e.g. \citealp{singh11}). Radio-loud AGN are popular candidates for unbiased studies of AGN because they can be selected using their low frequency extended radio lobe emission, which is known to be emitted isotropically, and is linked to the central AGN via the relativistic jets. Numerous studies have used radio-selected samples to probe the orientation dependence of AGN and test the orientation-based unified schemes (e.g. \citealp{jackson90}, \citealp{hes95}, \citealp{haas05}). 

We have been engaged in a major multi-wavelength study of a complete sample of radio selected, southern, 2Jy radio galaxies. In our previous work \citep{dicken09,dicken10} we investigated AGN power indicators in our samples and showed that [OIII] optical emission line luminosity ($L_{[\rm{OIII}]}$) is strongly correlated with both the mid-IR (24$\mu$m) and far-IR (70$\mu$m) continuum luminosities ($L_{24\mu m}$ and $L_{70\mu m}$ respectively). $L_{[\rm{OIII}]}$ is a potentially useful AGN power indicator because it is emitted far from the broad-line AGN nucleus (including the accretion disk component) and is therefore less likely to be affected by the obscuration due to dust structures in the near-nuclear regions (\citealp{tadhunter98}; \citealp{simpson98}). Therefore, the correlations between $L_{24\mu m}$, $L_{70\mu m}$ and $L_{[\rm{OIII}]}$ provide strong empirical evidence to support AGN illumination as the dominant heating mechanism for the thermal, mid- to far-IR (MFIR) emitting dust in powerful radio galaxies. However, this result is dependent on the reliability of the $L_{[\rm{OIII}]}$ as an AGN power indicator, in particular, whether the [OIII] line suffers some dust obscuration by the outer parts of the torus, or larger-scale dust structures in the host galaxies  (e.g. \citealp{jackson90}; \citealp{simpson98}). 

In addition, although the mid-IR continuum (20-30$\mu$m) emission is often assumed to be isotropic in nature and therefore  a good candidate for an AGN power indicator, there is some evidence in previous Spitzer and WISE investigations that obscuration could also affect the photometric flux measurements at mid-IR wavelengths (24$\mu$m) in radio galaxies (e.g. {\citealp{cleary07,leipski10,gurkan13}}). Moreover, the strong possibility of a starburst contribution   
rules out the longer wavelength, far-IR, continuum luminosities as reliable AGN power proxies \citep{dicken09,dicken12}. 

We further note that, although the low frequency radio emission is almost certainly isotropically emitted, the use of radio emission as an AGN power indicator has some notable drawbacks \citep{dicken08}. In particular, because the radio emission is emitted by lobes on large scales (up to hundreds of kpc),  it represents the jet power integrated over timescales of $\approx10^6$ years (e.g. \citealp{tadhunter12}). Moreover, the radio jet power is not necessarily closely coupled with AGN radiative power, because the amount of accretion energy converted into jet power may depend on parameters such as the spin of the supermassive black hole {(e.g. \citealp{blandford77, punsly90})}. In addition, radio luminosities are also potentially dependent on the gaseous environments with which the jets are interacting {\citep{barthel96}}, introducing variations in radio luminosities for otherwise identical intrinsic jet/AGN powers (\citealp{ramos13}; \citealp{hardcastle13}). 

Finally, although the X-ray continuum luminosity has the advantage that it is unequivocally linked to the AGN, due to its origin in the corona of the accretion disk, it suffers from the most variability compared to the other AGN power indicators. This is because it is emitted on the smallest scales, closest to the AGN, and therefore reflects better the instantaneous AGN power. In contrast, larger-scale emission components (e.g. the narrow line region) integrate the AGN power over longer timescales. Furthermore, careful modeling of the  X-ray observations is required to account for the effects of absorption and beamed non-thermal emission on the measured continuum fluxes (e.g. \citealp{hardcastle09}). 

Clearly, all the AGN power indicators have potential drawbacks. Therefore, it is important to investigate their
relative usefulness using complete samples of AGN with comprehensive multi-wavelength datasets.

This paper is the second in a series that analyses data from the Spitzer Infrared Spectrograph (IRS: \citealp{houck04};  program 50588: PI Tadhunter). The mid-IR spectral lines contained in these data provide important information about the degree of anisotropy, and hence utility of various AGN power indicators. For example, [OIV] $\lambda$25.89$\mu$m has a relatively high ionisation potential ($E_{ion}$=54.9eV) which favours an AGN origin; it also lies at the longer wavelength end of the IRS window,
and so is least likely to suffer from the effects of attenuation by the
circum-nuclear ISM. Indeed, several Spitzer spectroscopic studies have identified [OIV] $\lambda$25.89$\mu$m as a good candidate for an isotropic tracer of AGN power in samples of Seyfert galaxies (e.g. \citealp{diamond09}; \citealp{lamassa10}). 

Mid-IR spectral studies can also allow tests of the orientation-based unified schemes for AGN using measurements of the broad silicate features detected in absorption in some AGN and, relatively recently, also detected in emission in others (\citealp{siebenmorgen05}; \citealp{hao05}). 
Amorphous silicate grains have strong opacity peaks due to the Si--O stretching and the O--Si--O bending modes, leading to  features at 9.7$\mu$m and 18$\mu$m in the mid-IR spectrum. In a simple interpretation of the unified schemes, and as predicted by the theoretical work of \citet{nenkova02,nenkova08}, one may expect to detect silicate emission in AGN classified as broad-line objects at optical wavelengths in which circum-nuclear dust emission is relatively unattenuated. On the other hand, silicate absorption is likely to be associated with AGN classified as narrow-line objects at optical wavelengths, due to the attenuation of the mid-IR continuum by columns of dust along the line of sight. Many previous AGN studies have found this premise to hold in {\it general} terms (e.g. \citealp{shi06}; \citealp{hao07}; \citealp{gallimore10}; \citealp{landt10}). {However it is important to note that evidence for silicate emission has been detected in narrow-line quasars (\citealp{sturm06}, \citealp{nikutta09}) and type-2 Seyferts \citep{mason09}, which is explained in theoretical models by directly-illuminated dust within a clumpy torus \citep{nenkova08}.}

In the first paper of the series \citep{dicken12} we presented the spectra and a detailed description of the data analysis methodology, and also investigated the star formation properties of the samples. 
In this second paper we analyse the mid-IR emission line properties of the IRS spectra, focussing the study on diagnostics of AGN power, as well as testing the unified schemes for radio-loud AGN.  We assume a cosmology with $H_{o}=71$km s$^{-1} $Mpc$^{-1}, \Omega_{m}=0.27$ and $\Omega_{\lambda}=0.73$ throughout this paper.

\section{Samples and selection}
\label{sec:samples}
The 2Jy sample presented in this work is identical to that presented in our previous papers relating to the Spitzer Space Observatory observations of powerful southern radio galaxies (\citealp{tadhunter07}; \citealp{dicken08,dicken09,dicken10,dicken12}). It consists of all 46 powerful radio galaxies and steep-spectrum quasars ($ F_{\nu} \propto \nu^{-\alpha},\alpha^{4.8}_{2.7} > 0.5 $) selected from the 2Jy sample of \citet{wall85} with redshifts 0.05 $<z<$ 0.7, flux densities $S_{2.7\rm{GHz}}>$ 2~Jy and declinations $\delta<10^o$. 
The spectral index cut has been set to ensure that all the sources in the sample are dominated by steep spectrum lobe emission, while the lower redshift limit has been set to ensure that these galaxies are genuinely powerful radio sources. 

The mid-IR spectra analysed in this paper complement a wealth of data that has been obtained for the 2Jy sample over the last two decades. To date, these include: deep optical spectroscopic observations  (\citealp{tadhunter93, tadhunter98, tadhunter02}; \citealp{wills02}; \citealp{holt07}); extensive observations at radio wavelengths (\citealp{morganti93, morganti97, morganti99}; \citealp{dicken08}); complete deep optical imaging from Gemini \citep{ramos11a,ramos12}; deep Spitzer/MIPS and Herschel/PACS mid- to far-IR photometric observations (\citealp{dicken08}; Dicken et al. 2014 in prep.; detection rates 100, 90, 100 and  90\% at 24, 70, 100 and 160$\mu$m respectively). In addition, 98\% of the complete sample has recently been observed at X-ray wavelengths with XMM and Chandra \citep{mingo13}; and 78\% of the sample have deep 2.2 micron (K-band) near-infrared imaging \citep{inskip10}. 

A major advantage of our full sample of 46 objects is that it includes a range of optical broad-line radio galaxies and radio-loud quasars (BLRG/Q: 35\%), narrow-line
radio galaxies (NLRG: 43\%), and weak-line radio galaxies\footnote{WLRGs are defined as having low [OIII] emission line equivalent widths: EW([OIII]) $<$ 10\AA\, \citep{tadhunter98}. WLRG are also sometimes referred to as low excitation radio galaxies (LERG) in the literature (see \citealp{buttiglione09} for discussion).} (WLRG:
22\%). In terms of radio morphological classification, the sample
includes 72\% FRII sources, 13\% FRI sources, and 15\% compact steep
spectrum (CSS)/gigahertz peak spectrum (GPS) objects. The
sample is presented in Table \ref{tbl-1}\footnote{Additional information on the 2Jy sample can be found at http://2jy.extragalactic.info}.

We also present the results for a complete sub-sample of 17 3CRR radio-loud AGN\footnote{In \citet{dicken12} we stated that the 3CRR sample comprised of 19 objects, and that two objects overlap between the 3CRR and 2Jy samples (3C403 and 3C445). However, in fact, the latter two objects are drawn from the 3CR sample, rather than the  3CRR sample of Laing, Riley and Longair (1982). Because both of these objects are already in the 2Jy sample, and we do not consider the 2Jy and 3CRR samples separately in our analysis, none of our results are affected if we omit the two objects from the 3CRR sample.} selected from the sample of  \citet{laing83}. We have limited this sample to objects with FRII radio morphologies and redshifts $z$$\leq$0.1, leading to a sample which is complete in both Spitzer/MIPS detections (100\% at 24$\mu$m and 89\% at 70$\mu$m) and [OIII] $\lambda5007$ emission line flux measurements (100\%). The full sample of 17 objects also includes a range of BLRG/Q (12\%), NLRG (53\%), WLRG (35\%) class objects. Because the 3CRR objects have lower radio powers and redshifts on average than most of the 2Jy sample, they help to add statistical significance to the lower luminosity end of the 2Jy sample in our investigations (see \citet{dicken12} for more details).  Throughout the rest of this paper we will refer to this sample as the 3CRR sample. In addition, when discussing the investigation of the 2Jy and 3CRR sample objects together we will refer to this as the combined sample. 

\subsection{Data processing}

{The mid-IR spectra of 35 out of the 46 objects in the 2Jy sample were obtained in a dedicated campaign of Spitzer IRS observations (program 50558; P.I. Tadhunter) between July 2008 and March 2009 and first published in \citet{dicken12}. The spectra of 8 additional objects in the 2Jy sample and the 15 objects from the 3CRR sample were obtained from the Spitzer archive and observed between August 2004 and September 2006. These data came from various campaigns under several different PIs and, consequently, have varying integration times and observing modes. Note some of the latter data were first published in previous papers e.g. \citet{ogle06,leipski09}. }

Out of the 46 objects in the 2Jy sample, 43 (93\%) were successfully observed and detected by Spitzer/IRS. For the 17 3CRR sample objects 13 (76\%) were successfully observed and detected. The fully reduced rest-frame spectra for the 43 observed/detected objects in the 2Jy sample and the 13 objects from the 3CRR sample are presented in \cite{dicken12} and included here in an online appendix. \cite{dicken12} also present a detailed description of the observations and the data reduction methodology, which includes the use of the SMART (v.8.1.2.) program, developed by the IRS Team at Cornell University (\citealp{higdon04}; \citealp{lebouteiller10}). The spectra were optimally extracted using the full aperture of the IRS short low and long low slits. 

Since the publication of \citet{dicken12} there have been further updates to the data products from the Spitzer science centre. Slight improvements to the data handling for the pipeline processing and the data fitting are mentioned in the release notes for these new versions, for example the SMART team reports that the overall flux may be lower than previous calibrations by a few percent, due to slightly different stellar templates and a different photometric calibration.
We investigated the differences between latest data products and those presented in \citet{dicken12}, but found no significant differences. For example, the ratio of MIPS photometric flux to IRS flux at 24$\mu$m (measured over the response range of the MIPS filter) remained similar to the ratio ($MIPS_{24\mu m}/IRS_{24\mu m} \sim0.96$) published in \citet{dicken12}.

\section{Overview of spectra}
\label{sec:overview}

\begin{figure*}[t]
\epsscale{1.7}
\plotone{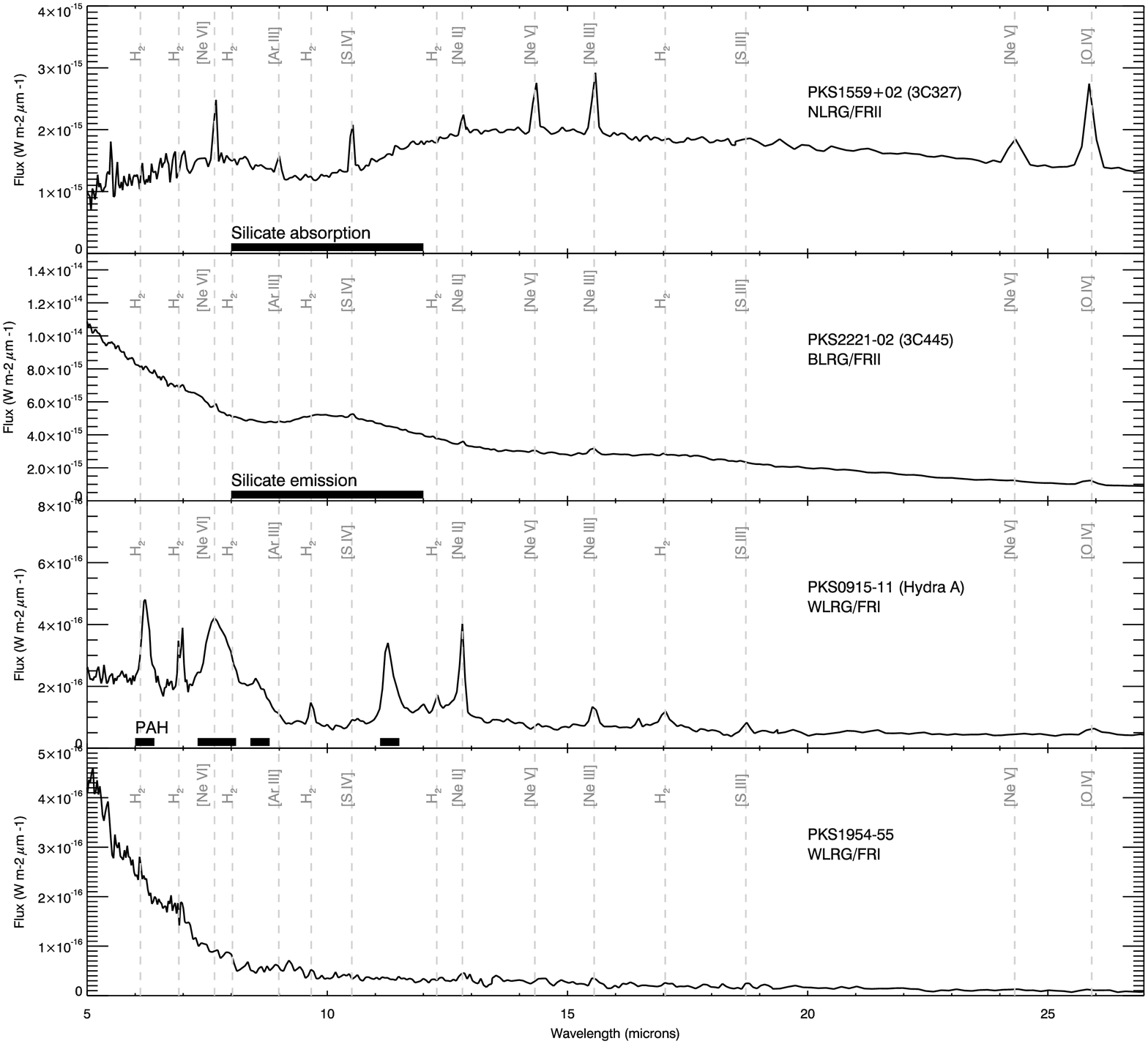}
\caption{Examples of IRS spectra from the 2Jy sample, presenting rest frame spectra from 5 to 27$\mu$m. From the top: PKS1559+02, shows strong mid-IR emission lines including the higher ionisation emission line ($E_{ion}$=54.9eV) [OIV] $\lambda$25.89$\mu$m as well as a silicate absorption at 10$\mu$m;  PKS2221-02, shows silicate emission at 10$\mu$m; PKS0915-11, shows strong lower ionisation lines e.g. [NeII] $\lambda$12.81$\mu$m  ($E_{ion}$=21.6eV) and $H_{2}$ (S(1) through S(6))  as well as strong PAH features at 6.2, 7.7, 8.6 and 11.3$\mu$m; PKS1954-55, shows a steep upturn in the spectrum at the blue end originating from the stellar continuum in the host galaxy.\label{fig:Ex_spec} }
\end{figure*}

The diversity in the mid-IR spectra of the 56 radio sources observed/detected in the 2Jy and 3CRR samples is striking (see \citealp{dicken12}, Figures \ref{spec1} to \ref{spec6} and the online appendix). However, many of the objects show features in common that are characteristic of AGN spectra at mid-IR wavelengths (see examples in Figure \ref{fig:Ex_spec}). 

Strong fine structure emission lines are detected in many of the spectra, with the most prominent being:
[ArII] $\lambda$ 6.96$\mu$m,
[Ne VI] $\lambda$ 7.65$\mu$m, 
[ArIII] $\lambda$8.99$\mu$m,
[SIV] $\lambda$10.51$\mu$m,
[NeII] $\lambda$12.81$\mu$m,
[Ne V] $\lambda$14.32$\mu$m,
[NeIII] $\lambda$15.56$\mu$m,
[SIII] $\lambda$18.71$\mu$m,
[Ne V] $\lambda$24.31$\mu$m and
 [OIV] $\lambda$25.89$\mu$m. 
Two objects (PKS2356-61, 3C33) have all of these emission lines detected in their spectra, whereas the majority of spectra have varying subsets of these lines detected. In some cases this may be due to the low S/N of the spectra but, alternatively, it may reflect the differing ionisation states of the emission line regions. It is noteworthy that all but one (PKS0035-02) of the 32 objects with strong optical emission lines (i.e. NLRG and BLRG/Q) have [OIV] $\lambda$25.89$\mu$m ($E_{ion}$=54.9eV) emission detected, when available in the rest frame spectral wavelength range. The detection of this line indicates a relatively high ionisation state, as expected given the presence of powerful AGN in many of the sources. In contrast, [OIV] is detected in only 4 out of the 12 observed/detected optical WLRG, perhaps reflecting the lower AGN bolometric luminosities of these objects \citep{dicken09}. 

Note that [OIV] $\lambda$25.89$\mu$m line may be blended with lower ionisation potential [FeII] $\lambda$25.99$\mu$m line ($E_{ion}$=7.9eV) in low resolution IRS spectra. However, high spectral resolution IRS studies of powerful radio galaxies show that the ratio of [OIV]/[FeII] is high for objects strong [OIV] detections \citep{guillard12}. It is not clear that the ratio is as high for weak [OIV] detections,  because the features are harder to de-blend when weak. Therefore, we consider it unlikely, but cannot entirely rule out, significant contamination of [OIV] fluxes by [FeII]. 

Many spectra are dominated by lower ionisation potential lines such as [NeII] $\lambda$12.81$\mu$m  ($E_{ion}$=21.6eV), which we have detected in 43/56 (77\%) of the sample,  and [NeIII] $\lambda$15.56$\mu$m ($E_{ion}$=41.0eV), detected in 49/56 (88\%). In addition, the high ionisation potential line [Ne VI] $\lambda$ 7.65$\mu$m ($E_{ion}$=99.1eV) is detected in 34\% (19/56) of the spectra, although generally at lower equivalent widths. We further note that $H_{2}$ lines (S(1) through S(7)), although not common in the samples overall, are detected in 10/56 objects (18\%); this is discussed further in Section \ref{sec:H2}. {Previously, \citet{ogle10} detected strong $H_{2}$ emission lines  in 31\% of 3C and 3CRR radio galaxies at z$<$0.22. These radio molecular hydrogen emission galaxies (radio MOHEGs) are thought to be powered by jet-shocked molecular gas. }

The overall shape of the mid-IR continuum can be significantly affected by prominent 10$\mu$m silicate emission or absorption features, as discussed in Section \ref{sec:intro}. These features are now well documented (e.g. \citealp{shi06}) and are sometimes associated with an additional, less prominent, feature at 18$\mu$m. However, the 18$\mu$m feature is not detected in absorption in any of the IRS spectra for the objects in our samples (see Section \ref{sec:silicate}).

Broad, starburst tracing, PAH features at 7.7 and 11.3$\mu$m are also detected in 9 out of the 56 objects with good IRS spectra. In addition, low equivalent width 11.3$\mu$m PAH features are detected in a further 8 objects. Other PAH bands that can make a significant contribution to the spectral continuum shape in the mid-IR are the 8.6$\mu$m band and a blend of PAH emission at 17$\mu$m (17.38 \& 17.87$\mu$m), although the latter can be strongly contaminated by $H_2$(S1) emission. 

Finally, the contribution of starlight from the host galaxies can also affect the shape of the IRS spectra at short wavelengths ($\lambda < 8$$\mu$m).  This effect is seen as a sharp upturn in flux at the blue end of the spectra, and is particularly apparent for objects with low power AGN, for example, the WLRG PKS1839-48 and PKS1954-55.  

\begin{deluxetable}{c@{\hspace{0mm}}c@{\hspace{0mm}}r@{\hspace{0mm}} 
c@{\hspace{0mm}}r@{\hspace{0mm}}c@{\hspace{0mm}}r@{\hspace{0mm}}
c@{\hspace{0mm}}r@{\hspace{0mm}}c@{\hspace{0mm}}r@{\hspace{0mm}}
c@{\hspace{0mm}}r@{\hspace{0mm}}c@{\hspace{0mm}}r@{\hspace{0mm}}
c@{\hspace{0mm}}r@{\hspace{0mm}}c@{\hspace{0mm}}r@{\hspace{0mm}}c@{\hspace{0mm}}
}
\tabletypesize{\scriptsize}
\tablecaption{2Jy Sample - Mid-IR Line Fluxes\label{tbl-1}}
\tablewidth{0pt}
\tablehead{
\colhead{PKS Name}{\hspace{-1mm}} &\colhead{$z$}{\hspace{3mm}} & \colhead{[ArIII]}{\hspace{-3mm}} 
& \colhead{}{\hspace{-1mm}} &\colhead{[NeVI]}{\hspace{-3mm}} & \colhead{}{\hspace{-1mm}} &\colhead{[SIV]}{\hspace{-3mm}}
&\colhead{}{\hspace{-1mm}} &\colhead{[NeII]}{\hspace{-3mm}} & \colhead{}{\hspace{-1mm}}&\colhead{[NeV]}{\hspace{-3mm}} 
&\colhead{}{\hspace{-1mm}} & \colhead{[NeIII]}{\hspace{-3mm}} &\colhead{}{\hspace{-1mm}}& \colhead{[SIII]}{\hspace{-3mm}} 
&\colhead{}{\hspace{0mm}} & \colhead{[NeV]}{\hspace{-3mm}} & \colhead{}{\hspace{0mm}} &\colhead{[OIV]}{\hspace{-3mm}}
& \colhead{}{\hspace{0mm}} 
}
\startdata
$\lambda$ ($\mu m$)	&			&		6.985	&			&		7.652	&			&		10.511	&			&		12.814	&			&		14.322	&			&		15.555	&			&		18.713	&			&		24.318	&			&		25.890	&			\\
$E_{ion}$ (eV)	&			&		15.8	&			&		99.1	&			&		34.8	&			&		21.6	&			&		97.1	&			&		41.0	&			&		23.3	&			&		97.1	&			&		54.9	&			\\
\cutinhead{}																																																										\\
0023$-$26	&	\phantom{a}	0.322	&		5.7	&	$\pm$	0.2	&		--	&			&		--	&			&		15.9	&	$\pm$	0.5	&		--	&			&		8.9	&	$\pm$	2.7	&		--	&			&		--	&			&		4.5	&	$\pm$	1.2	\\
0034$-$01	&	\phantom{a}	0.073	&		--	&			&		--	&			&		--	&			&		3.6	&	$\pm$	0.5	&		--	&			&		3.1	&	$\pm$	1.6	&		--	&			&		--	&			&	$<$	2.7	&			\\
0035$-$02	&	\phantom{a}	0.220	&		ááá 	&			&		--	&			&		--	&			&		7.5	&	$\pm$	0.2	&		--	&			&		8.4	&	$\pm$	0.7	&		--	&			&		--	&			&	$<$	4.9	&			\\
0038$+$09	&	\phantom{a}	0.188	&		6.8	&	$\pm$	0.1	&		4.5	&	$\pm$	0.5	&		--	&			&		12.1	&	$\pm$	0.6	&		3.6	&	$\pm$	0.6	&		12.5	&	$\pm$	4.6	&		--	&			&		--	&			&		11.4	&	$\pm$	1.2	\\
0039$-$44	&	\phantom{a}	0.346	&		3.6	&	$\pm$	0.1	&		17.5	&	$\pm$	0.7	&		--	&			&		16.8	&	$\pm$	2.2	&		26.4	&	$\pm$	2.1	&		46.6	&	$\pm$	3.1	&		17.1	&	$\pm$	0.6	&		23.8	&	$\pm$	2.2	&		91.0	&	$\pm$	3.0	\\
0043$-$42	&	\phantom{a}	0.116	&		--	&			&		--	&			&		--	&			&	$<$	17.3	&			&		--	&			&		4.7	&	$\pm$	0.8	&		--	&			&		--	&			&	$<$	2.2	&			\\
0105$-$16	&	\phantom{a}	0.400	&		--	&			&		--	&			&		--	&			&	$<$	7.9	&			&		--	&			&	$<$	5.1	&			&		--	&			&		--	&			&		or	&			\\
0117$-$15	&	\phantom{a}	0.565	&		--	&			&		--	&			&		--	&			&		4.2	&	$\pm$	0.3	&		--	&			&		5.4	&	$\pm$	0.5	&		--	&			&		or	&			&		or	&			\\
0213$-$13	&	\phantom{a}	0.147	&		--	&			&		11.8	&	$\pm$	0.4	&		9.4	&	$\pm$	0.3	&		9.5	&	$\pm$	0.9	&		7.1	&	$\pm$	0.5	&		13.3	&	$\pm$	2.1	&		--	&			&		9.8	&	$\pm$	0.5	&		29.6	&	$\pm$	5.0	\\
0235$-$19	&	\phantom{a}	0.620	&		--	&			&		--	&			&		--	&			&		8.1	&	$\pm$	2.0	&		--	&			&		10.3	&	$\pm$	0.3	&		--	&			&		or	&			&		or	&			\\
0252$-$71	&	\phantom{a}	0.566	&		--	&			&		--	&			&		--	&			&		5.1	&	$\pm$	0.2	&		--	&			&	$<$	4.3	&	$\pm$	0.6	&		--	&			&		or	&			&		or	&			\\
0347$+$05	&	\phantom{a}	0.339	&		1.8	&	$\pm$	0.1	&		--	&			&		--	&			&		10.6	&	$\pm$	1.8	&		--	&			&		4.1	&	$\pm$	1.3	&		--	&			&		--	&			&	$<$	3.4	&			\\
0349$-$27	&	\phantom{a}	0.066	&		--	&			&		--	&			&		--	&			&		9.6	&	$\pm$	0.3	&		--	&			&		15.6	&	$\pm$	0.5	&		--	&			&		--	&			&		15.8	&	$\pm$	6.5	\\
0404$+$03	&	\phantom{a}	0.089	&		--	&			&		--	&			&		--	&			&		13.6	&	$\pm$	0.4	&		13.9	&	$\pm$	0.5	&		18.4	&	$\pm$	0.6	&		--	&			&		--	&			&		18.8	&	$\pm$	0.6	\\
0409$-$75	&	\phantom{a}	0.693	&		--	&			&		--	&			&		--	&			&		7.6	&	$\pm$	1.4	&		--	&			&		3.8	&	$\pm$	0.6	&		--	&			&		or	&			&		or	&			\\
0442$-$28	&	\phantom{a}	0.147	&		--	&			&		--	&			&		--	&			&		4.3	&	$\pm$	0.1	&		--	&			&		6.5	&	$\pm$	2.3	&		--	&			&		--	&			&		7.4	&	$\pm$	0.2	\\
0620$-$52	&	\phantom{a}	0.051	&		--	&			&		--	&			&		--	&			&		2.0	&	$\pm$	0.1	&		--	&			&		2.1	&	$\pm$	0.3	&		--	&			&		--	&			&		2.2	&	$\pm$	0.5	\\
0625$-$35	&	\phantom{a}	0.055	&		--	&			&		--	&			&		--	&			&		3.4	&	$\pm$	0.1	&		--	&			&	$<$	5.7	&			&		--	&			&		--	&			&	$<$	2.1	&			\\
0625$-$53	&	\phantom{a}	0.054	&		--	&			&		--	&			&		--	&			&	$<$	2.7	&			&		--	&			&	$<$	3.8	&			&		--	&			&		--	&			&	$<$	2.6	&			\\
0806$-$10	&	\phantom{a}	0.110	&		--	&			&		98.8	&	$\pm$	3.3	&		112.1	&	$\pm$	3.7	&		38.1	&	$\pm$	1.3	&		90.2	&	$\pm$	5.2	&		145.4	&	$\pm$	21.6	&		--	&			&		117.7	&	$\pm$	3.9	&		365.2	&	$\pm$	29.4	\\
0859$-$25	&	\phantom{a}	0.305	&		--	&			&		--	&			&		6.0	&	$\pm$	0.3	&		3.5	&	$\pm$	0.6	&		--	&			&		11.5	&	$\pm$	0.5	&		--	&			&		--	&			&		14.1	&	$\pm$	3.2	\\
0915$-$11	&	\phantom{a}	0.054	&		14.0	&	$\pm$	0.5	&		--	&			&		--	&			&		33.3	&	$\pm$	1.1	&		--	&			&		14.3	&	$\pm$	0.8	&		5.2	&	$\pm$	0.6	&		--	&			&		5.1	&	$\pm$	2.4	\\
0945$+$07	&	\phantom{a}	0.086	&		--	&			&		--	&			&		--	&			&		10.3	&	$\pm$	1.0	&		--	&			&		5.4	&	$\pm$	0.2	&		--	&			&		--	&			&		22.4	&	$\pm$	0.7	\\
1136$-$13	&	\phantom{a}	0.554	&		--	&			&		7.7	&	$\pm$	0.3	&		22.7	&	$\pm$	2.9	&	$<$	11.7	&			&		--	&			&		11.0	&	$\pm$	4.9	&		--	&			&		or	&			&		or	&			\\
1151$-$34	&	\phantom{a}	0.258	&		4.2	&	$\pm$	0.1	&		5.2	&	$\pm$	0.5	&		5.2	&	$\pm$	0.2	&		6.3	&	$\pm$	1.2	&		--	&			&		11.0	&	$\pm$	1.2	&		6.6	&	$\pm$	0.7	&		--	&			&		12.4	&	$\pm$	3.4	\\
1306$-$09	&	\phantom{a}	0.464	&		--	&			&		--	&			&		--	&			&	$<$	8.8	&			&		--	&			&	$<$	13.2	&			&		--	&			&		or	&			&		or	&			\\
1355$-$41	&	\phantom{a}	0.313	&		--	&			&		13.8	&	$\pm$	1.6	&		--	&			&	$<$	8.4	&			&		15.8	&	$\pm$	0.8	&	$<$	19.0	&			&		--	&			&		19.4	&	$\pm$	5.3	&		51.5	&	$\pm$	1.7	\\
1547$-$79	&	\phantom{a}	0.483	&		--	&			&		13.0	&	$\pm$	1.2	&		23.9	&	$\pm$	4.6	&	$<$	11.7	&			&		--	&			&		14.0	&	$\pm$	1.6	&		--	&			&		or	&			&		or	&			\\
1559$+$02	&	\phantom{a}	0.104	&		--	&			&		120.3	&	$\pm$	4.0	&		103.4	&	$\pm$	3.4	&		41.5	&	$\pm$	1.4	&		111.5	&	$\pm$	3.7	&		141.3	&	$\pm$	4.7	&		--	&			&		146.8	&	$\pm$	6.6	&		378.7	&	$\pm$	17.7	\\
1602$+$01	&	\phantom{a}	0.462	&		--	&			&		--	&			&		--	&			&		6.3	&	$\pm$	0.6	&		--	&			&	$<$	5.3	&			&		--	&			&		or	&			&		or	&			\\
1648$+$05*	&	\phantom{a}	0.154	&		--	&			&		--	&			&		--	&			&		--	&			&		--	&			&		--	&			&		--	&			&		--	&			&		--	&			\\
1733$-$56	&	\phantom{a}	0.098	&		--	&			&		--	&			&		6.4	&	$\pm$	0.2	&		17.9	&	$\pm$	0.6	&		--	&			&		17.3	&	$\pm$	0.6	&		10.8	&	$\pm$	0.8	&		--	&			&		16.7	&	$\pm$	0.6	\\
1814$-$63	&	\phantom{a}	0.063	&		--	&			&		10.8	&	$\pm$	0.4	&		--	&			&		77.4	&	$\pm$	2.6	&		--	&			&		62.8	&	$\pm$	2.1	&		--	&			&		--	&			&		32.2	&	$\pm$	2.5	\\
1839$-$48	&	\phantom{a}	0.112	&		--	&			&		--	&			&		--	&			&	$<$	1.5	&			&		--	&			&	$<$	30.2	&			&		--	&			&		--	&			&	$<$	1.7	&			\\
1932$-$46	&	\phantom{a}	0.231	&		--	&			&		--	&			&		--	&			&		5.7	&	$\pm$	1.5	&		--	&			&		8.6	&	$\pm$	1.2	&		--	&			&		--	&			&		10.9	&	$\pm$	0.7	\\
1934$-$63	&	\phantom{a}	0.183	&		4.2	&	$\pm$	0.5	&		3.1	&	$\pm$	0.1	&		--	&			&		18.5	&	$\pm$	2.2	&		--	&			&		14.1	&	$\pm$	0.5	&		7.6	&	$\pm$	0.3	&		--	&			&		8.4	&	$\pm$	2.7	\\
1938$-$15	&	\phantom{a}	0.452	&		--	&			&		--	&			&		--	&			&		6.7	&	$\pm$	4.6	&		--	&			&		5.2	&	$\pm$	1.7	&		--	&			&		--	&			&		or	&			\\
1949$+$02	&	\phantom{a}	0.059	&		--	&			&		--	&			&		50.7	&	$\pm$	1.7	&		30.7	&	$\pm$	1.0	&		29.5	&	$\pm$	1.0	&		58.7	&	$\pm$	1.9	&		--	&			&		--	&			&		156.7	&	$\pm$	5.2	\\
1954$-$55	&	\phantom{a}	0.060	&		--	&			&		--	&			&		--	&			&		2.6	&	$\pm$	0.3	&		--	&			&		2.5	&	$\pm$	0.1	&		--	&			&		--	&			&	$<$	1.6	&			\\
2135$-$14*	&	\phantom{a}	0.200	&		--	&			&		--	&			&		--	&			&		--	&			&		--	&			&		--	&			&		--	&			&		--	&			&		--	&			\\
2135$-$20	&	\phantom{a}	0.635	&		6.6	&	$\pm$	0.5	&		--	&			&	(	12.1	)&	$\pm$	2.8	&		42.8	&	$\pm$	0.4	&		--	&			&		23.7	&	$\pm$	8.4	&		8.6	&	$\pm$	1.6	&		or	&			&		or	&			\\
2211$-$17*	&	\phantom{a}	0.153	&		--	&			&		--	&			&		--	&			&		--	&			&		--	&			&		--	&			&		--	&			&		--	&			&		--	&			\\
2221$-$02	&	\phantom{a}	0.057	&		--	&			&		2.2	&	$\pm$	0.1	&		289.6	&	$\pm$	9.6	&		38.4	&	$\pm$	2.8	&		--	&			&		99.0	&	$\pm$	5.8	&		--	&			&		--	&			&		105.0	&	$\pm$	4.7	\\
2250$-$41	&	\phantom{a}	0.310	&		--	&			&		--	&			&		9.6	&	$\pm$	0.3	&		8.4	&	$\pm$	0.4	&		7.1	&	$\pm$	1.3	&		13.2	&	$\pm$	1.7	&		3.8	&	$\pm$	0.1	&		7.5	&	$\pm$	1.0	&		30.7	&	$\pm$	1.0	\\
2314$+$03	&	\phantom{a}	0.220	&		12.1	&	$\pm$	0.4	&		--	&			&		--	&			&		60.8	&	$\pm$	3.5	&		--	&			&		20.0	&	$\pm$	0.7	&		--	&			&		--	&			&		26.8	&	$\pm$	7.4	\\
2356$-$61	&	\phantom{a}	0.096	&		7.0	&	$\pm$	0.2	&		16.2	&	$\pm$	0.5	&		2.8	&	$\pm$	0.5	&		18.9	&	$\pm$	0.6	&		11.3	&	$\pm$	0.8	&		19.8	&	$\pm$	3.8	&		13.2	&	$\pm$	0.4	&		22.8	&	$\pm$	0.8	&		55.0	&	$\pm$	1.8	\\
\enddata
\tablecomments{Table presenting prominent mid-IR line fluxes for the 2Jy sample. Units of $10^{-18}$ Watts $m^{-2}$. *PKS1648 was not detected by Spitzer IRS in our program -- PKS2135-14 and PKS2211-17 were not observed as their mid-IR fluxes were deemed to low for detection.}  
\end{deluxetable}

\begin{deluxetable}{c@{\hspace{0mm}}c@{\hspace{0mm}}r@{\hspace{0mm}}
c@{\hspace{0mm}}r@{\hspace{0mm}}c@{\hspace{0mm}}r@{\hspace{0mm}}
c@{\hspace{0mm}}r@{\hspace{0mm}}c@{\hspace{0mm}}r@{\hspace{0mm}}
c@{\hspace{0mm}}r@{\hspace{0mm}}c@{\hspace{0mm}}r@{\hspace{0mm}}
c@{\hspace{0mm}}r@{\hspace{0mm}}c@{\hspace{0mm}}r@{\hspace{0mm}}c@{\hspace{0mm}}
}
\tabletypesize{\scriptsize}
\tablecaption{3CRR Sample - Mid-IR Line Fluxes\label{tbl-2}}
\tablewidth{0pt}
\tablehead{
\colhead{Name}{\hspace{-1mm}} &\colhead{z}{\hspace{3mm}} & \colhead{[ArIII]}{\hspace{-3mm}} 
& \colhead{}{\hspace{-1mm}} &\colhead{[NeVI]}{\hspace{-3mm}} & \colhead{}{\hspace{-1mm}} &\colhead{[SIV]}{\hspace{-3mm}}
&\colhead{}{\hspace{-1mm}} &\colhead{[NeII]}{\hspace{-3mm}} & \colhead{}{\hspace{-1mm}}&\colhead{[NeV]}{\hspace{-3mm}} 
&\colhead{}{\hspace{-1mm}} & \colhead{[NeIII]}{\hspace{-3mm}} &\colhead{}{\hspace{-1mm}}& \colhead{[SIII]}{\hspace{-3mm}} 
&\colhead{}{\hspace{0mm}} & \colhead{[NeV]}{\hspace{-3mm}} & \colhead{}{\hspace{0mm}} &\colhead{[OIV]}{\hspace{-3mm}}
& \colhead{}{\hspace{0mm}} 
}

\startdata
$\lambda$ ($\mu m$)	&		&		6.985	&			&		7.652	&			&		10.511	&			&		12.814	&			&		14.322	&			&		15.555	&			&		18.713	&			&		24.318	&			&		25.890	&			\\
$E_{ion}$ (eV)	&		&		15.8	&			&		99.1	&			&		34.8	&			&		21.6	&			&		97.1	&			&		41.0	&			&		23.3	&			&		97.1	&			&		54.9	&			\\
\cutinhead{}																																																									\\
3C33	&	0.060	&		13.4	&	$\pm$	0.4	&		27.9	&	$\pm$	0.9	&		16.3	&	$\pm$	0.5	&		33.9	&	$\pm$	1.1	&		14.4	&	$\pm$	0.5	&		48.7	&	$\pm$	1.6	&		27.3	&	$\pm$	0.9	&		13.4	&	$\pm$	1.4	&		72.9	&	$\pm$	2.4	\\
3C35*	&	0.067	&		--	&			&		--	&			&		--	&			&		--	&			&		--	&			&		--	&			&		--	&			&		--	&			&		--	&			\\
3C98	&	0.030	&		--	&			&		--	&			&		--	&			&		10.1	&	$\pm$	1.0	&		--	&			&		23.3	&	$\pm$	0.8	&		--	&			&		--	&			&		44.2	&	$\pm$	1.5	\\
3C192	&	0.060	&		--	&			&		--	&			&		--	&			&		6.2	&	$\pm$	0.4	&		--	&			&		4.8	&	$\pm$	1.7	&		6.0	&	$\pm$	2.4	&		--	&			&		22.8	&	$\pm$	0.8	\\
3C236	&	0.101	&		--	&			&		--	&			&		--	&			&		19.9	&	$\pm$	2.0	&		--	&			&	$<$	21.9	&			&		--	&			&		--	&			&	$<$	11.0	&			\\
3C277.3*	&	0.085	&		--	&			&		--	&			&		--	&			&		--	&			&		--	&			&		--	&			&		--	&			&		--	&			&		--	&			\\
3C285	&	0.079	&		--	&			&		--	&			&		15.0	&	$\pm$	0.5	&		18.5	&	$\pm$	0.6	&		--	&			&		34.2	&	$\pm$	1.1	&		--	&			&		15.3	&	$\pm$	0.5	&		56.8	&	$\pm$	1.9	\\
3C293	&	0.045	&		29.9	&	$\pm$	1.0	&		--	&			&		--	&			&		73.2	&	$\pm$	2.4	&		--	&			&		30.5	&	$\pm$	1.0	&		9.0	&	$\pm$	1.6	&		--	&			&		5.6	&	$\pm$	2.0	\\
3C305	&	0.042	&		--	&			&		--	&			&		51.2	&	$\pm$	1.7	&		42.9	&	$\pm$	1.4	&		27.5	&	$\pm$	0.9	&		147.9	&	$\pm$	4.9	&		36.7	&	$\pm$	1.2	&		35.1	&	$\pm$	1.2	&		189.6	&	$\pm$	6.3	\\
3C321	&	0.096	&		--	&			&		114.0	&	$\pm$	3.8	&		137.0	&			&		26.2	&	$\pm$	0.9	&		140.3	&	$\pm$	9.9	&		210.5	&	$\pm$	6.9	&		18.0	&	$\pm$	6.9	&		207.2	&	$\pm$	9.9	&		736.6	&	$\pm$	32.9	\\
3C326	&	0.090	&		--	&			&		--	&			&		--	&			&	$<$	2.8	&			&		--	&			&	$<$	59.2	&			&		--	&			&		--	&			&	$<$	3.3	&			\\
3C382	&	0.058	&		0.6	&	$\pm$	0.1	&		71.5	&	$\pm$	2.4	&		88.8	&	$\pm$	2.9	&		28.4	&	$\pm$	0.9	&		9.6	&	$\pm$	0.6	&		46.3	&	$\pm$	1.5	&		--	&			&		11.2	&	$\pm$	1.7	&		30.5	&	$\pm$	1.0	\\
3C388	&	0.092	&		--	&			&		--	&			&		--	&			&	$<$	2.8	&			&		--	&			&	$<$	4.9	&			&		--	&			&		--	&			&		2.4	&	$\pm$	0.6	\\
3C390.3	&	0.056	&		13.1	&	$\pm$	0.4	&		8.0	&	$\pm$	1.5	&		17.2	&	$\pm$	1.2	&		42.0	&	$\pm$	1.4	&		--	&			&		30.6	&	$\pm$	1.0	&		--	&			&		--	&			&		25.5	&	$\pm$	0.8	\\
3C452	&	0.081	&		7.8	&	$\pm$	1.7	&		4.0	&	$\pm$	0.8	&		2.7	&	$\pm$	0.8	&		19.4	&	$\pm$	1.1	&		--	&			&		21.0	&	$\pm$	3.4	&		6.0	&	$\pm$	1.2	&		--	&			&		11.8	&	$\pm$	0.4	\\
4C73.08*	&	0.058	&		--	&			&		--	&			&		--	&			&		--	&			&		--	&			&		--	&			&		--	&			&		--	&			&		--	&			\\
da240*	&	0.036	&		--	&			&		--	&			&		--	&			&		--	&			&		--	&			&		--	&			&		--	&			&		--	&			&		--	&			\\
\enddata

\tablecomments{Table presenting prominent mid-IR line fluxes for the 3CRR sample. Units of $10^{-18}$ W/$m^2$. *3C35 and 3C277.3 were not detected in the observations taken in mapping mode, and 3C4C73.08 and da240 do not have IRS data available in the Spitzer Archive.  }  
\end{deluxetable}

\subsection{Fitting the spectra}
\label{sec:lines}

In order to fit the spectra we used PAHFIT v1.2\footnote{PAHFIT is made available under the terms of the GNU General Public License.}, which is an IDL program developed by J.D.T Smith and B.T. Draine for studying PAH features in the mid-IR spectra of the inner regions of local star-forming galaxies \citep{smith07}. The fitted spectra for all objects can be seen in the online appendix. We experimented with, and adapted, the PAHFIT code to check and improve its performance for the samples of powerful radio-loud AGN presented in this study. Firstly, we experimented with adding dust components into the model with black body temperatures 400K, 600K, 1000K  and 2000K, in order to account for the potential hotter dust continuum features due to the powerful AGN, but the PAHFIT model fitted the data just as well without these extra hot components. Therefore, we returned the code to the default model continuum dust temperatures (300K, 135K, 90K, 65K, 50K, 40K, 35K). However, it is notable that, in fitting the blue end of the spectrum where the hot dust components would be, the PAHFIT program adds a stellar continuum component, which could be compensating for the lack of hot dust components in the model instead of tracing the stellar continuum. Because we are not directly interested in the hot dust or stellar contribution in this paper, the fit to the continuum seen in the PAHFIT results is adequate for our purposes. 

Secondly, when fitting the 10$\mu$m and 18$\mu$m silicate absorption features, we found that the default PAHFIT extinction model, based on the dust opacity law of \cite{kemper04}, did not fit the features well. After some experimentation it was found that the depth of the silicate absorption feature was better fitted in models assuming high ratios of 10$\mu$m to 18$\mu$m opacity. This implies that the 18$\mu$m absorption feature is weak for our samples of powerful radio-loud AGN, as we discuss further in Section \ref{sec:silicate}. 

Thirdly, by default PAHFIT includes PAH features  in the model fitted to the spectra. However, in \citet{dicken12} we found little direct evidence for PAH features and/or the starburst activity often associated with such emission, in the vast majority of objects. Therefore, in this investigation, when fitting the spectra with PAHFIT to measure the fluxes of the fine structure lines, we only included PAH features in the model for objects with confirmed detections of the 11.2$\mu$m feature (see Table 5  in \citealp{dicken12} ). It is notable that the PAHFIT model without the PAH components still fits the observed spectra well in the objects without evidence for PAH. 

Finally, the original PAHFIT model does not include silicate emission, because objects with such emission were not included in the sample that PAHFIT was originally designed to fit \citep{smith07}. However, the 10$\mu$m silicate emission feature is clearly detected at varying strengths in 25\% of the observed/detected objects in the 2Jy and 3CRR samples. Therefore we made a further adaptation of the  PAHFIT program to fit silicate emission as well as absorption, following the method employed in \citet{gallimore10}. This entailed adding a model for warm dust clouds that are optically thin at infrared wavelengths. This assumes that the clouds have a simple, slab geometry and opacity at 10$\mu$m, $\tau_{10} < 1$. These warm clouds are further assumed to be partially covered by cold, absorbing dust clouds, and the source function is a scaled Planck spectrum at  the fitted temperature for simplicity. 

The PAHFIT measurements of the most prominent, mid-IR emission lines are presented in Tables \ref{tbl-1} and \ref{tbl-2}, while the $H_2$ detections and silicate emission/absorption strengths are presented in Tables \ref{tbl-3} and \ref{tbl-4}. 

PAHFIT returns formal statistical uncertainties on the emission line fits. However, given the wide range of factors affecting the quality of these measurements, we found these formal uncertainties to be unrealistically small in many cases. In addition, PAHFIT may return an emission line measurement when in fact it has just fitted a particular line to the fluctuations in the noise. Therefore, we adopted a more robust approach to the emission line detection and uncertainty calculations.

In determining whether a particular feature was detected we first visually inspected all the spectra for candidate detections. Following this we re-fitted the data for each of the two nod positions of the observations and measured the difference in emission line flux between these two spectra. If the difference in the line flux between the two nod position measurements was less than 20\% we considered this a confident detection. Objects that had differences between 20\% and 50\% were re-confirmed or rejected by visual inspection. Note that the signal to noise improves in the combined nodded spectra, which adds weight to these weaker detections. The uncertainties presented are the differences between the mean flux in the averaged nod position spectra and the fluxes in the individually nodded spectra, the calibration uncertainty of 3.3\%\footnote{Note that the flux calibration is anchored at 24$\mu$m, with an uncertainty of the order 3\% (Sloan. G. C., private communication).  Therefore, toward shorter wavelengths the uncertainty will increase by several percent because the spectrum moves through several orders and changes module once.}, or the PAHFIT uncertainty, whichever was greater. Note the uncertainties for the spectra that were obtained in Mapping mode may be underestimated because they do not have nodded data for comparison. These lines were identified visually and the quoted error is either the calibration uncertainty or the PAHFIT uncertainty, whichever is greater. 

Also, in Tables \ref{tbl-1} and \ref{tbl-2} we present upper limits for the main emission lines ([NeII] $\lambda$12.81$\mu$m, [NeIII] $\lambda$15.56$\mu$m, [OIV] $\lambda$25.89$\mu$m) used in this investigation, for objects in which these lines were not clearly detected. These upper limits were derived by using, as a template, the IRS spectrum of PKS2356-61 which has clearly detected narrow emission lines and little evidence for contaminating PAH emission. After subtracting the silicate absorption fitted in the PAHFIT model the spectrum of PKS2356-61 was added to the spectrum of each object without detected emission lines. A multiplicative scaling factor for the template spectrum was then varied until the appropriate emission line feature was just detected in a visual inspection of the combined spectrum. Multiplying the measured emission line flux for PKS2356-61 by the scaling factor then gives a robust upper limit on emission line flux. For high redshift objects the [NeII] $\lambda$12.81$\mu$m line is redshifted from the SL module to the LL module. Therefore, for consistency, we used the spectrum of the higher redshift object PKS2250-41 to measure the upper limits for sources where this line was observed in the LL module rather than SL.

\section{AGN power indicators}
\label{sec:AGNpower}

The mid-IR spectral line [OIV] $\lambda$25.89$\mu$m is a good candidate for an AGN power indicator because it has a relatively high ionisation potential ($E_{ion}$=54.9eV) which favours an AGN rather than a starburst origin. This feature lies towards the longer wavelength end of the IRS spectral range, and so is less likely to suffer from the effects of dust extinction. We note that, in the case of the 2Jy and 3CRR samples, the mid-IR [OIV] $\lambda$25.89$\mu$m line is detected in all but one of the objects with strong optical emission lines for which [OIV] lies in the rest frame wavelength range of the IRS instrument ($z<$0.350).

To investigate whether the [OIV] line emission indeed has an AGN origin for the 2Jy and 3CRR objects, we compare it to other AGN power indicators from our extensive archive of complementary observations at optical, infrared, radio and X-ray wavelengths\footnote{The Spitzer/MIPS, [OIII] and 5GHz luminosities used in the following investigation are presented in \citet{dicken09} and \citet{dicken10} for the 2Jy and 3CRR samples respectively, whereas [OIII] fluxes for the 3CRR sample were taken from \citet{buttiglione09}. X-ray luminosities were taken from the study of \citet{mingo13}}. 

\begin{figure*}
\epsscale{1.9}
\plotone{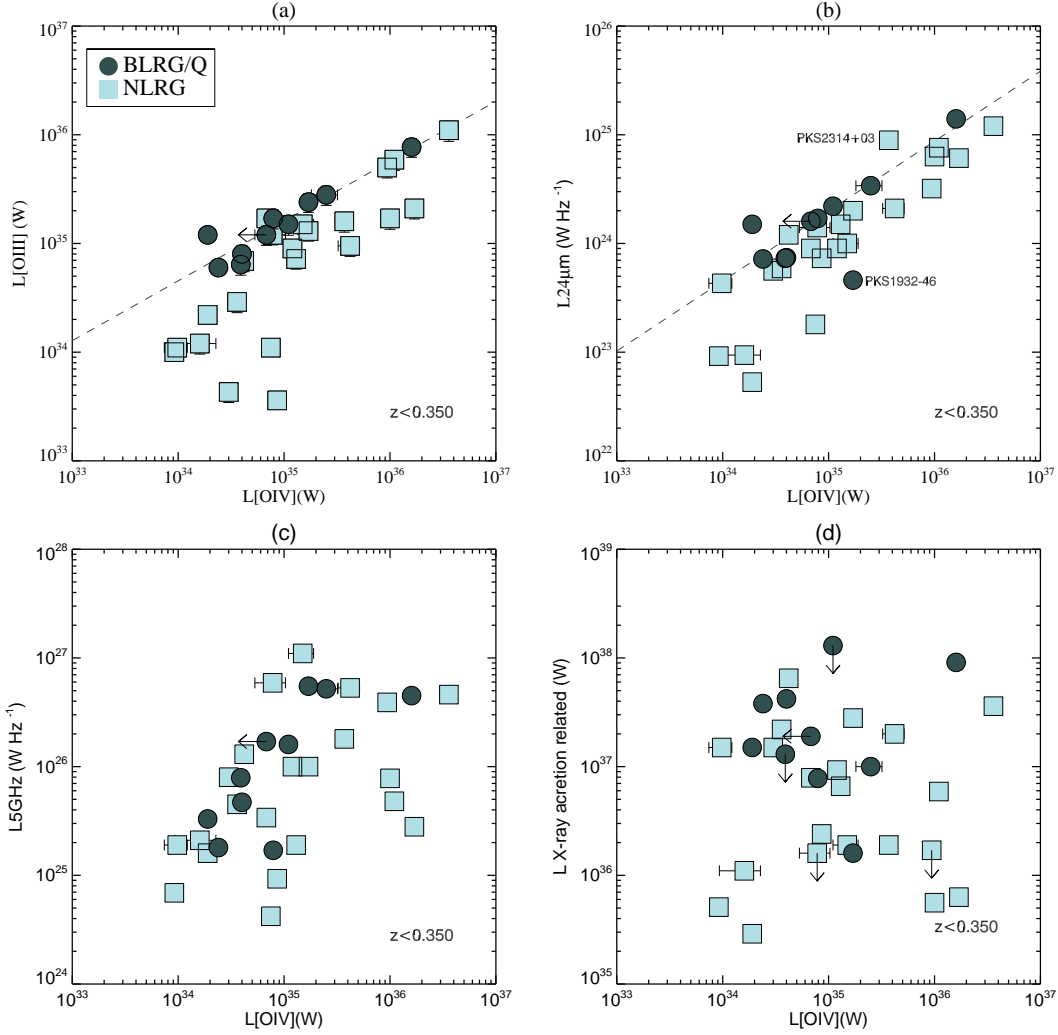}
\caption{Plot showing $L_{[\rm{OIV}]}\lambda$25.89$\mu$m versus $L_{[\rm{OIII}]}\lambda5007$, $L_{24\mu m}$, $L_{5GHz}$  and $L_{\rm{X-ray}}$ with objects identified by optical classes. The samples are limited to objects with $z<$0.350 due to the requirement that the [OIV] line line in the redshift range of Spitzer/IRS. In order to maintain a comparable range in the x and y-axes, plot (d) excludes the object 3C305 which has a low upper limit value $L_{\rm{X-ray}}$ = $<8.9\times10^{33}$. Optical classifications are: BLRG/Q --- broad-line radio galaxy/quasar; NLRG --- narrow-line radio galaxy. Uncertainties are plotted for all data points but are typically smaller than the object marker. The regression lines fitted in plots (a) and (b) are fitted to the BLRG/Q objects. See Section \ref{sec:opt} for discussion of PKS1932-46 and Section \ref{sec:shape} for discussion of PKS2314+03. \label{fig:O4plot} }
\end{figure*}

For the initial analysis, we do not include the objects classified as WLRG at optical wavelengths. Mounting evidence suggests that these objects are fundamentally different from optical strong-line radio galaxies (e.g. \citealp{hardcastle07}; \citealp{buttiglione10}). In particular, they have been associated with different accretion modes for the central super-massive black holes \citep[e.g.][]{hardcastle07}. Moreover, the detection rate of [OIV] for the WLRG in the two samples is low (only 2/12 observed/detected objects). This is to be expected if  WLRG  have intrinsically weak AGN {(e.g. \citealp{ogle06,dicken09})}, so that the higher ionisation lines, such as [OIV], are less likely to be detected (see Section \ref{sec:weak} for further discussion).

Figure \ref{fig:O4plot} plots $L_{[\rm{OIV}]}$ against four other candidate AGN power indicators; [OIII] $\lambda5007$ optical emission line luminosity ($L_{[\rm{OIII}]}$\footnote{$L_{[\rm{OIII}]}$ values are corrected for Galactic extinction.}), mid-infrared luminosity at 24$\mu$m ($L_{24\mu m}$), total radio luminosity at 5GHz ($L_{5GHz}$) and accretion related X-ray luminosity ($L_{\rm{X-ray}}$\footnote{We use the X-ray accretion-related component here as a proxy for AGN power, rather than the X-ray unabsorbed component, which 
is likely to be affected by beaming, because it is emitted by the jet. The accretion related X-ray luminosity has been integrated over energies 2-10 keV assuming an intrinsic power-law continuum shape  (see \citet{mingo13} for further details.)} for the combined sample. This figure only includes objects with redshifts $z<$0.350 for which the [OIV] line is available in the wavelength range of Spitzer IRS. This excludes 11 objects from the higher redshift range of the 2Jy sample. 

$L_{[\rm{OIII}]}$, $L_{24\mu m}$, $L_{5GHz}$ and $L_{\rm{X-ray}}$ are widely accepted to have an AGN origin in radio galaxies: optical [OIII] emission arises from narrow-line region clouds that are directly heated by the AGN. In the absence of strong star formation, the 24$\mu$m continuum emission originates from warm/hot dust situated relatively close to, and heated directly by, the AGN. Low-frequency radio lobe emission is known to be isotropic and is powered by the relativistic jets originating from the AGN, here we use the complete data set of total 5Ghz radio fluxes that are available for all the sample objects.  This is a relatively high radio frequency band, but due to the steep spectrum selection and the careful analysis detailed in \citet{dicken08} we are certain that the potential effect of beaming at these higher frequencies is minimal for the strong-line objects in the 2Jy sample. Lastly, the accretion-related X-ray emission originates from, or close to, the accretion disk associated with the AGN.

\begin{deluxetable}{l@{\hspace{0mm}}c@{\hspace{0mm}}c@{\hspace{0mm}}c@{\hspace{0mm}}c@{\hspace{0mm}}c}
\tabletypesize{\scriptsize}
\tablecaption{Combined Sample Spearman Rank Statistics \label{tbl-3}}
\tablewidth{0pt}
\tablehead{ \colhead{Rank Correlation}{\hspace{0mm}} &\colhead{n}{\hspace{0mm}} &
\colhead{$r_s$}{\hspace{0mm}} &\colhead{significance}{\hspace{0mm}}  &
\colhead{$r_s$ (ASURV)}{\hspace{0mm}} &\colhead{significance}{\hspace{0mm}}  }

\startdata 		
\cutinhead{Basic rank correlation statistics}
 $L_{[\rm{OIV}]}$ vs $L_{[\rm{OIII}]}$ 	&	32	&	0.79	&	$>99.9$\%	&	0.78	&	$>99.9$\%	\\
 $L_{[\rm{OIV}]}$ vs $L_{24}$ 	&	32	&	0.81	&	$>99.9$\%	&	0.79	&	$>99.9$\%	\\
 $L_{[\rm{OIV}]}$ vs $L_{5GHz}$ 	&	32	&	0.55	&	$>99.9$\%	&	0.53	&	$>99.9$\%	\\
 $L_{[\rm{OIV}]}$ vs $L_{\rm{X-ray}}$ 	&	32	&	-0.01	&	$\ll 50$\%	&	-0.003  &	$\ll 50$\% 	\\
 $L_{[\rm{NeIII}]}$ vs $L_{[\rm{OIII}]}$ 	&	44	&	--	&	--	&	0.75	&	$>99.9$\%	\\
 $L_{[\rm{NeIII}]}$ vs $L_{24}$ 	&	44	&	--	&	--	&	0.76	&	$>99.9$\%	\\
 $L_{[\rm{OIII}]}$ vs $L_{24}$ 	&	44	&	--	&	--	&	0.86	&	$>99.9$\%	\\
\cutinhead{Partial rank correlation with \emph{$z$}}											
 $L_{[\rm{OIV}]}$ vs $L_{[\rm{OIII}]}$ 	&	32	&	0.63	&	$>99.9$\%	&	0.62	&	$>99.9$\%	\\
 $L_{[\rm{OIV}]}$ vs $L_{24}$ 	&	32	&	0.67	&	$>99.9$\%	&	0.65	&	$>99.9$\%	\\
 $L_{[\rm{OIV}]}$ vs $L_{5GHz}$ 	&	32	&	-0.24	&	$\ll 50$\%	&	-0.22	&	$\ll 80$\%	\\
 $L_{[\rm{OIV}]}$ vs $L_{\rm{X-ray}}$ 	&	32	&	-0.27	&	$\ll 50$\%	&	0.17  &	$\ll 50$\%	\\
\enddata 

\tablecomments{Results of various Spearman rank correlation tests for the combined 3CRR
and 2Jy sample for the correlations presented in Figure \ref{fig:O4plot}. These tests do not include weak-line radio galaxies, but do include one object (PKS0035-02) with an [OIV] upper limit value and the five objects with X-ray upper limits which are included in the test as if they were detections. Values of $0<r_s<1$ are given for each correlation, where a
value close to 1 is highly significant. Columns 2 and 3 present the results of the spearman rank statistics; note the single upper limit in $L_{[\rm{OIV}]}$ values was treated as a detections for the purposes of these test.  Columns 4 and 5 present statistics for all the objects in the combined sample, handling the upper limits using survival analysis statistics.}
\end{deluxetable}

From Figure \ref{fig:O4plot} it is clear that $L_{[\rm{OIII}]}$ and $L_{24\mu m}$  are strongly correlated with $L_{[\rm{OIV}]}$, whereas the correlations are much weaker for $L_{5GHz}$ and $L_{\rm{X-ray}}$. 
Spearman rank statistics for the four correlations shown in Figure \ref{fig:O4plot} are presented in Table \ref{tbl-3}. The top section of the table presents the basic rank statistics, which shows that all the AGN power proxies correlate significantly with $L_{[\rm{OIV}]}$ apart from X-ray luminosities; the null hypothesis that the variables are unrelated is rejected at the $>$99.5\% confidence level.  The correlations between $L_{[\rm{OIII}]}$, $L_{24\mu m}$ and $L_{5GHz}$, which are common indicators of AGN power, give weight to the hypothesis that $L_{[\rm{OIV}]}$ is also a reliable AGN power indicator. 

Although the majority of the basic correlation statistics support the use of $L_{[\rm{OIV}]}$ as an AGN power proxy, it is important to test whether the correlations are intrinsic, and not due to the increase in luminosity with redshift in the flux-limited samples. Therefore, in the second section of Table \ref{tbl-3}, we present Spearman partial rank correlation tests. These tests are aimed at determining whether the correlations only arise because the variables are independently correlated with a third variable, in this case redshift. The results of the partial correlation tests show that we can still reject the null hypothesis that the variables are unrelated at the $>$99.5\% level of significance, for the correlations  with $L_{[\rm{OIII}]}$ and $L_{24\mu m}$, but not for the $L_{[\rm{OIV}]}$ vs $L_{5GHz}$ correlation. In the case of $L_{\rm{X-ray}}$ we find no statistically significant correlation with $L_{[\rm{OIV}]}$ in our investigation.

The null result for the partial rank $L_{[\rm{OIV}]}$ versus $L_{5GHz}$ correlation test reflects the scatter seen in Figure \ref{fig:O4plot}(c) and agrees with our previous work \citep{dicken09}, where we found that the correlation between $L_{[\rm{OIII}]}$ and $L_{24\mu m}$ has greater significance than the correlation between $L_{5GHz}$ and $L_{24\mu m}$. As mentioned in the introduction, the lack of correlation with $L_{5GHz}$ may be due to the bulk of the radio emission originating hundreds of kpc from the nucleus, introducing a time lag between the recent power level of the AGN (as traced by $L_{[\rm{OIV}]}$) and that of the radio emission. Alternatively, the luminosity of the radio emission could also depend on secondary factors, other than intrinsic AGN power, such as the properties of the ISM in the halo of the host galaxy into which the jets and lobes expand \citep{barthel96}. It is also noteworthy that \citet{mingo13} found a better correlation between radio luminosity and jet-related X-ray luminosity. 

Statistically significant correlations between $L_{[\rm{OIII}]}$ and $L_{\rm{X-ray}}$ were seen in the studies of \citet{hardcastle09} and \citet{mingo13}. For the sample included in this paper we find no significant correlation between $L_{[\rm{OIII}]}$ and $L_{\rm{X-ray}}$ ($r_s$ = 0.12, confidence level 50\%). The difference between our results and those of \citet{hardcastle09} and \citet{mingo13} can be explained in terms of the large intrinsic scatter for the correlations. This large apparent scatter in $L_{\rm{X-ray}}$  may be caused by variability in the X-ray source that is not seen for [OIII] and 24$\mu$m, because these are emitted on larger scales and therefore integrate the power of the AGN over longer time scales. In the case of the XMM observations used for the higher redshift objects, it is possible that the X-ray emission of the target objects is contaminated by neighbouring sources or emission from hot X-ray haloes of the host galaxies/clusters. Finally, uncertainties in the correction for the gas absorption, and separating the non-thermal X-ray emission from the jets, might also contribute to the scatter. We note that a similarly larger scatter in the correlations involving $L_{\rm{X-ray}}$ and $L_{5GHz}$ were also seen in the analysis of intrinsic AGN power indicators in Seyfert galaxies by \citet{lamassa10}. Our sample is also significantly smaller and covers a more restricted range in redshift/luminosity. However, even with the smaller sample, $L_{[\rm{OIV}]}$, $L_{[\rm{OIII}]}$ and $L_{24\mu m}$ still show significant correlations in our analysis. Therefore we conclude that the latter are better bolometric indicators of AGN power than $L_{\rm{X-ray}}$. 

As a comparison to our statistical method, and to take into account the effect of the one $L_{[\rm{OIV}]}$ upper limit for PKS0035-02  included in the previous tests as a detection, as well as the 5 upper limits in $L_{\rm{X-ray}}$, the above Spearman rank tests were also conducted using survival statistics in the ASURV package (\citealp{isobe86}; \citealp{lavalley92}), as implemented in IRAF. The results of the ASURV Spearman rank tests are presented in columns 4 and 5 of Table \ref{tbl-3} and show results that are consistent with those of the basic tests, albeit with slightly lowered $r_s$ values. 

\begin{figure*}
\epsscale{2}
\plotone{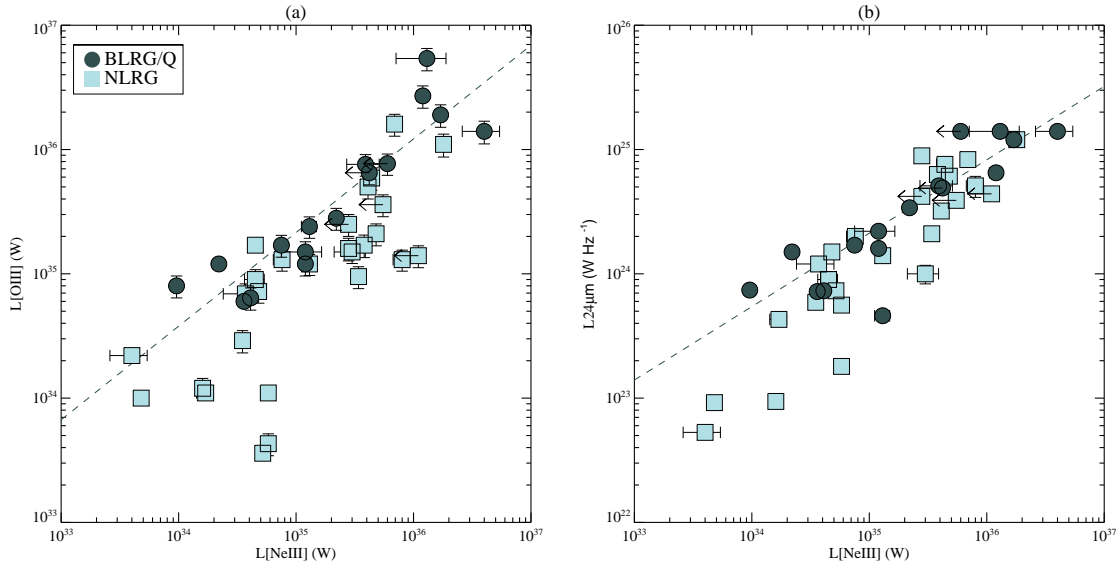}
\caption{Plot showing $L_{[\rm{NeIII}]}\lambda$15.56$\mu$m versus $L_{[\rm{OIII}]}\lambda5007$  and $L_{24\mu m}$ with objects identified by optical classes, see Figure \ref{fig:O4plot} for definitions. The linear regression results for the BLRG/Q objects
are shown as dashed lines. \label{fig:Ne3plot} }
\end{figure*}

A disadvantage of using the [OIV] $\lambda$25.89$\mu$m line is that it is redshifted out of the spectral range of the IRS instrument for objects with redshifts $z$$\geq$0.35. This affects 11/46 objects in the higher redshift half of the 2Jy sample. Therefore, we also investigate correlations 
involving the [NeIII] $\lambda$15.56$\mu$m line, which also has a relatively high ionisation energy ($E_{ion}$=41.0eV) and may also have an AGN origin, following similar arguments to those presented above for [OIV]. However, [NeIII] has a lower ionisation energy than [OIV] and therefore is more likely
to have a significant contribution from stellar photoionisation; also, its shorter wavelength may make it more susceptible to obscuration. 

To test the value of [NeIII] $\lambda$15.56$\mu$m as an alternative AGN power proxy, we have plotted $L_{[\rm{NeIII}]}$ against $L_{[\rm{OIII}]}$  and $L_{24\mu m}$ in Figure \ref{fig:Ne3plot}. Strong correlations between these quantities can clearly be seen in Figures \ref{fig:Ne3plot}(a) and (b), and the correlations are confirmed with the ASURV\footnote{We only use ASURV statistics in this analysis due to the 5 upper limits on the [NeIII] emission. } Spearman rank statistical tests presented in Table \ref{tbl-3}. These show that we can reject the null hypothesis that the variables are unrelated at a $>$99.5\% level. However, we note that the correlations between  $L_{[\rm{NeIII}]}$, $L_{[\rm{OIII}]}$  and $L_{24\mu m}$ have lower statistical significance  than those between  $L_{[\rm{OIV}]}$,$L_{[\rm{OIII}]}$  and $L_{24\mu m}$, despite the larger sample ($n = 44$ and 34 for the tests with [NeIII] and [OIV] respectively). We conclude that [NeIII] $\lambda$15.56$\mu$m is a good AGN proxy for powerful radio galaxies, however less so than [OIV] $\lambda$25.89$\mu$m, {as further supported by the results presented in Section \ref{sec:opt}. }

\section{Degree of anisotropy}
\label{sec:opt}

Previously, in  \citet{dicken09,dicken10}, we investigated the isotropy of the optical [OIII] emission in the 2Jy sample. Despite a large scatter in the results, we found evidence that broad-line radio galaxies and radio-loud quasars (BLRG/Q) showed a tendency to fall below the $L_{[\rm{OIII}]}$ vs $L_{70\mu m}$ correlation, but on the  $L_{[\rm{OIII}]}$ vs $L_{24\mu m}$ correlation defined by the narrow-line radio galaxies (NLRG). In agreement with the orientation-based unified schemes, this result can be explained if both the [OIII] and 24$\mu$m emission, but not the 70$\mu m$ emission, in NLRG suffer a mild degree of dust extinction (e.g. by a dust torus), however, this previous work was not statistically conclusive. We now return to this investigation with the new mid-IR emission line data.


The optical classifications of the combined sample objects are indicated in Figure \ref{fig:O4plot} and \ref{fig:Ne3plot}. From a detailed inspection of Figure \ref{fig:O4plot}(a) it is clear that the BLRG/Q are offset from the NLRG by typically 0.2 -- 1.0 dex, tending to the upper edge of the correlation between $L_{[\rm{OIV}]}$ and $L_{[\rm{OIII}]}$.  In Figure \ref{fig:O4plot}(b) we find a similar offset is also apparent for the $L_{[\rm{OIV}]}$ versus $L_{24\mu m}$ correlation\footnote{\label{foot:1932} We note that the outlier BLRG that lies under the correlation of Figures \ref{fig:O4plot}(b), \ref{fig:Ne3plot}(b)  is PKS1932-46. This object has been uniquely identified with an under-luminous AGN that is proposed to have just switched to a low activity phase, see \citet{inskip07} for details.}. Figures \ref{fig:Ne3plot}(a) \& (b) also show a similar result for the correlations between $L_{[\rm{NeIII}]}$, $L_{[\rm{OIII}]}$ and $L_{24\mu m}$. In addition, there is evidence that the difference is larger for objects with lower AGN powers: below $L_{[\rm{OIV}]}$ = $10^{35}$~W there is a group of NLRG objects that lie $\sim$1.5 to 2 dex below the correlation defined by the BLRG/Q.

The [OIV] and [NeIII] lines and the 24$\mu m$ continuum are emitted within the same mid-IR wavelength range. Therefore, naively one might expect that any obscuration that affects the  24$\mu m$ continuum will also affect the [OIV] and [NeIII] lines to a similar degree. However, whereas much of the 24$\mu$m continuum emission is thought to be emitted by the inner parts of the torus ($<$10~pc) and therefore could potentially suffer significant dust extinction in the NLRG objects, even at these relatively long wavelengths, the [OIV] and [NeIII] are likely to be emitted on the much larger scales of the NLR (10 -- 1,000 pc) for which the level of dust extinction is relatively modest in the mid-IR (but more significant, of course, for optical emission lines such as [OIII] $\lambda5007$).

\begin{figure*}[t]
\epsscale{1.8}
\plotone{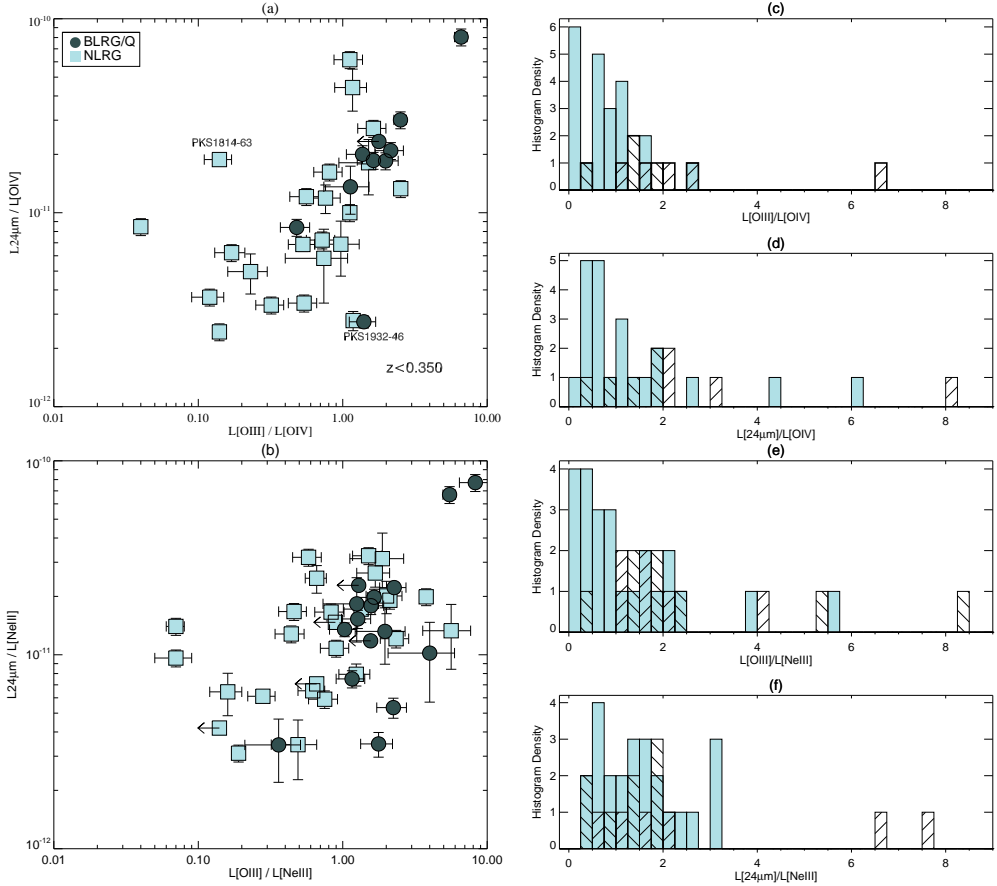}	
\caption{Plots of diagnostic ratios [OIII] $\lambda$5007/[OIV] $\lambda$25.89$\mu$m and 24$\mu$m/[OIV] $\lambda$25.89$\mu$m as well as [OIII] $\lambda$5007/[NeIII] $\lambda$15.56$\mu$m and 24$\mu$m/[NeIII] $\lambda$15.56$\mu$m.  The upper limit in [OIV] for PKS0035-02 is plotted on plot (a) but not on plot (c) and (d). The five upper limits in [NeIII] results are plotted in plot (b) but not in plot (e) and (f). Ratio value presented in plot (d) and (f) are in units of $10^{11}$. See Figure \ref{fig:O4plot} for plot symbol definitions. Crossed bars in the histograms are BLRG/Q and filled bars are NLRG. \label{fig:Op_diag} }
\end{figure*}

To visualise the evidence for anisotropy in another way we can study the ratios of [OIII]/[OIV]  and 24$\mu m$/[OIV] as well as [OIII]/[NeIII] and 24$\mu m$/[NeIII]. Assuming that the [OIV] and [NeIII] line are emitted isotropically, dividing the [OIII] and 24$\mu m$ emission by the mid-IR lines serves to highlight any difference between the optical classes of objects due to obscuration. Figure \ref{fig:Op_diag} plots [OIII]/[OIV] vs 24$\mu m$/[OIV], where a loose correlation with some scatter can be seen. A caveat to this investigation is that the [OIII]/[OIV] ratio may be affected by ionisation effects that also increase the scatter. As expected from Figure \ref{fig:O4plot}(a) and (b) the BLRG/Q have, on average, higher ratios of [OIII]/[OIV]  and 24$\mu m$/[OIV]  than the NLRG, and the scatter is greater for the NLRG than for the BLRG/Q\footnote{The NLRG outlier above the correlation, PKS1814-63, is a rare example of radio source in a disc galaxy \citep{morganti11}. It is possible that star formation in the disk could be contributing to its 24$\mu$m emission.}. Figure \ref{fig:Op_diag}(a) confirms that the same NLRG that have evidence for attenuation of $L_{[\rm{OIII}]}$ compared to BLRG/Q also have evidence for attenuation of $L_{24\mu m}$, suggesting that both types of emission are significantly attenuated in these objects. We have also plotted the ratios as histograms in Figures \ref{fig:Op_diag}(c),(d) separating BLRG/Q and NLRG. The degree of attenuation of the [OIII] and 24$\mu$m emission can be quantified using the median ratio values for the two optical classes: [OIII]/[OIV] = 1.6(BLRG/Q) and 0.7(NLRG); and 24$\mu$m/[OIV] = 1.9(BLRG/Q) and 0.8(NLRG); both sets of numbers imply a factor $\sim$2.3 attenuation in the NLRG relative to the BLRG/Q objects. 

Although the degree of attenuation appears to be to the same degree for [OIII] and 24$\mu$m and in broadly the same objects, it is clear that a factor of 2.3 attenuation at 5007\AA\, implies a much lower visual dust extinction ($A_V$) than a similar factor at 24$\mu m$. For example, to attenuate the [OIII] by a factor of 2.3 we would require $A_V$$\sim$0.8 visual extinction compared to $A_V$$\sim$20-100 (depending on the extinction law; \citealp{mathis90}; \citealp{draine03}) implied by the 24$\mu m$ results. Therefore, the absorbing dust columns must be different in the two cases. It is also the case that, if the [OIV] is emitted by the same region as the [OIII], it will suffer lower extinction ($A_V$$\sim$0.1). This supports our assumption that the [OIV] is unattenuated. 

\begin{figure}[t]
\epsscale{0.8}
\plotone{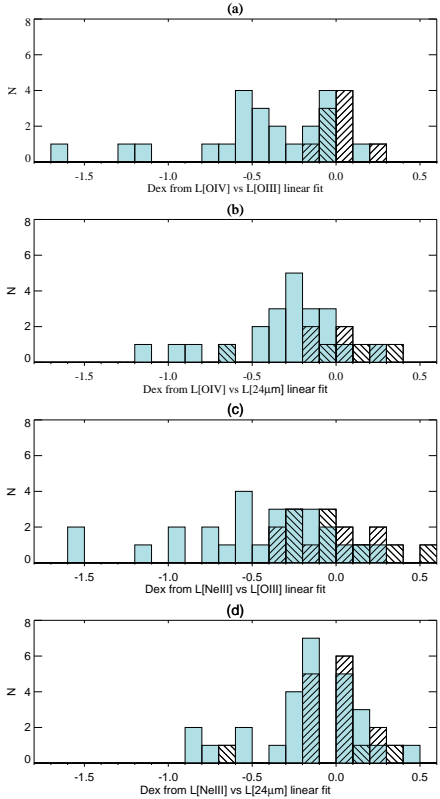}	
\caption{Plots showing the displacement of the sample objects from a regression line fitted to the BLRG/Q for the data presented in Figure \ref{fig:O4plot}(a) \& (b) and Figure \ref{fig:Ne3plot}(a) \& (b)\label{fig:hist_dex}. Crossed bars in the histograms are BLRG/Q and filled bars are NLRG. The results are shown in dex. }
\end{figure}

To put these results from Figure \ref{fig:O4plot}(a) \& (b) on a firmer statistical footing, we have considered the vertical displacements of the BLRG and NLRG objects from a linear regression line fitted to the BLRG/Q objects alone. The results are presented in Figure \ref{fig:hist_dex}, where we show histograms of the distribution of displacements for the BLRG/Q and NLRG. Using a 1D KS test we can reject the null hypothesis that the BLRG/Q and NLRG samples are drawn from the same population at better than a 
99\% significance level for the $L_{[\rm{OIV}]}$ vs $L_{[\rm{OIII}]}$ and $L_{[\rm{OIV}]}$ vs $L_{24\mu m}$ correlations\footnote{The test were conducted with and without the 1 upper limit in [OIV] and the 5 upper limits in [NeIII] with no significant change in the results}. Using the regression line fitted to the BLRG/Q objects in Figure \ref{fig:O4plot}(a) as a reference, we find that the NLRG lie a median factor of 2.7 (mean factor 5.5$\pm$0.3)  below the BLRG/Q in  $L_{[\rm{OIII}]}$ for a given AGN power, as traced by $L_{[\rm{OIV}]}$. The ranges of attenuation factors derived using this technique are consistent with those derived above using the line ratios.


To include the 11 objects higher redshift objects that have [OIV] emitted outside of the spectral range of Spitzer IRS, we repeat the investigation in Figure \ref{fig:Op_diag}(b), replacing [OIV] with [NeIII]. From the histograms in Figures \ref{fig:Op_diag}(e),(f), the line ratios relative to [NeIII] show a difference for [OIII], but not for the the 24$\mu$m emission, where the median ratio values are: [OIII]/[NeIII] = 1.7(BLRG/Q) and 0.8(NLRG); and 24$\mu$m/[NeIII] = 1.4(BLRG/Q) and 1.4(NLRG). 
For the correlations between $L_{[\rm{NeIII}]}$ vs $L_{[\rm{OIII}]}$ the 1D KS  tests show we can reject the null hypothesis that the BLRG/Q and NLRG samples are drawn from the same population at no better than a 90\% level of significance. Testing the  $L_{[\rm{NeIII}]}$ vs $L_{24\mu m}$ correlation we find that we can reject the null hypothesis at the $<$90\% level of significance, both for tests with and without PKS1932-46. These tests confirm what is readily identifiable by eye: that the offset between the BLRG/Q and NLRG in Figure  \ref{fig:Ne3plot} is not as significant as seen in Figure \ref{fig:O4plot}(a) \& (b). Plausibly, [NeIII] emission is subject to a greater degree of extinction than the longer wavelength [OIV] emission, and photoionisation by stars may make a significant contribution to the [NeIII] emission (as discussed in Section \ref{sec:star}); these factors are likely to increase the scatter in the correlation plots. Again using a regression line fitted to the BLRG/Q objects in Figure \ref{fig:O4plot} as a reference, we find that the NLRG have [OIII] luminosities that are a median factor of 2.2 (mean factor 5.4$\pm$1.2) lower,  and $L_{24\mu m}$  that are a median factor of 1.4 (mean factor1.9$\pm$1.2) lower, than the BLRG/Q for a given AGN power traced by $L_{[\rm{NeIII}]}$. 

Three objects are easily identifiable in Figure \ref{fig:O4plot}(a) with $L_{[\rm{OIV}]} <10^{35}$~W as a group lying well below the correlation defined by the NLRG, with factors of 13, 20 and 41 lower $L_{[\rm{OIII}]}$ than expected for a BLRG/Q of equivalent $L_{[\rm{OIV}]}$. One possible explanation for this result is that large-scale disc structures in the host galaxies cause enhanced dust extinction of the [OIII] emission. It is notable that the host galaxies of all of these outlier objects show evidence for large-scale disks with dust lanes (3C285 - \citealp{madrid06}, 3C305 - \citealp{chiaberge99}; PKS1814-63 - \citealp{morganti11}); they are also amongst the objects with the strongest $H_{2}$ detections. This $H_{2}$ emission could plausibly originate in the same dust disk described above, adding weight to this hypothesis.

\subsection{Mid-infrared $H_2$ emission lines}
\label{sec:H2}

\begin{deluxetable}{c@{\hspace{1mm}}
c@{\hspace{0mm}}c@{\hspace{0mm}}r@{\hspace{0mm}}c@{\hspace{0mm}}r@{\hspace{0mm}}c@{\hspace{0mm}}r
}
\tabletypesize{\scriptsize}
\tablecaption{$H_2$ Detections \label{tbl-4}}
\tablewidth{0pt}
\tablehead{
\colhead{Name}{\hspace{-1mm}} & \colhead{$z$}{\hspace{-1mm}} 
& \colhead{$H_2 (S3)$}{\hspace{-1mm}} & \colhead{}{\hspace{-1mm}} &\colhead{$H_2 (S2)$}{\hspace{-1mm}} & \colhead{}{\hspace{-1mm}} & \colhead{$H_2 (S1)$} & \colhead{}{\hspace{-1mm}} 
}

\startdata
0023$-$26	&	0.322	&		4.0	&	$\pm$	0.1	&		5.2	&	$\pm$	4.1	&		10.4	&	$\pm$	1.8	\\
0915$-$11	&	0.054	&		16.3	&	$\pm$	1.2	&		6.3	&	$\pm$	0.2	&		9.1	&	$\pm$	1.9	\\
1151$-$34	&	0.258	&		5.3	&	$\pm$	0.2	&		2.5	&	$\pm$	0.9	&		4.1	&	$\pm$	0.1	\\
1733$-$56	&	0.098	&		3.7	&	$\pm$	0.1	&		0.5	&	$\pm$	0.2	&		10.3	&	$\pm$	0.4	\\
1814$-$63	&	0.063	&		7.6	&	$\pm$	0.3	&		8.7	&	$\pm$	0.3	&		18.2	&	$\pm$	6.1	\\
3C33	&	0.060	&	$<$	ul	&			&	$<$	ul	&			&		9.2	&	$\pm$	1.1	\\
3C285	&	0.079	&		8.4	&	$\pm$	0.3	&		12.5	&	$\pm$	0.4	&		26.5	&	$\pm$	0.0	\\
3C293	&	0.045	&		97.6	&	$\pm$	3.2	&		22.7	&	$\pm$	0.7	&		74.8	&	$\pm$	3.6	\\
3C305	&	0.042	&		27.9	&	$\pm$	0.9	&		11.9	&	$\pm$	0.4	&		33.1	&	$\pm$	1.1	\\
3C326	&	0.090	&	$<$	ul	&			&	$<$	ul	&			&		7.4	&	$\pm$	2.8	\\
\enddata

\tablecomments{Table: Units of $10^{-18}$ W/$m^2$. *3C285 observed in mapping mode, therefore uncertainties may be underestimated, see Section \ref{sec:lines}.}  
\end{deluxetable}

{Comprehensive studies of $H_2$ in radio galaxies are presented in \citet{ogle07,ogle10,nesvadba10,guillard12}. These studies have shown that radio-loud molecular hydrogen emission galaxies (radio MOHEGs) have extremely high $H_{2}$/PAH ratios, indicative of jet shock-excitation of $H_{2}$. We have detected $H_{2}$ lines ((S(1) through S(7)) in a minority of objects (10/56, 18\%). Such $H_{2}$ lines are an indicator of shocked molecular gas as well as a tracer of the molecular gas contents of the host galaxies.  It is notable that of the 2Jy sample the $H_{2}$ detections in the IRS spectra occur only in objects that also have PAH detections \citep{dicken12}. 

The fact that $H_2$ and PAH are detected together suggests a link between the two emission features. Indeed, given that PAH features are known indicators of young stellar populations (YSP),  the $H_2$ features may be tracing the total reservoir of molecular gas that fuels recent episodes of star formation in the host galaxies. However, aside from the detection statistics, we do not find a strong correlation between PAH and $H_2$ equivalent widths for objects in which both features are detected. This may reflect the fact that the strength of the $H_2$  emission depends not only on the total reservoir of molecular gas,
but also on how the molecular emission is excited. Indeed, \citet{ogle10,nesvadba10} have shown that the $H_2$ emission is likely to be excited by shocks dissipating mechanical energy in at least two of our samples objects (3C293, 3C326). It is also important to bear in mind that PAH emission itself is not necessarily a good quantitative indicator of the star formation rates \citep{dicken12}. }

We further note that 8 of the 10 of the objects with $H_2$ detections have relatively low redshifts z$<$0.1, and the remaining two objects are both compact steep spectrum objects which may tend towards denser/richer circum-nuclear gas environments than more extended radio galaxies \citep{tadhunter11}. Because it is unlikely that we are seeing an evolutionary effect over this relatively narrow redshift range, the $H_2$ detections could show a preference to occur in radio galaxies that have large-scale disks; such objects are more common at lower redshifts and radio powers in the two samples. Clear examples of such disk-like radio galaxies include PKS1814-63, 3C293 and 3C305 (referred to in the previous section). 

\begin{deluxetable}{c@{\hspace{0mm}}c@{\hspace{0mm}}c@{\hspace{0mm}}c@{\hspace{0mm}}r@{\hspace{5mm}}c@{\hspace{1mm}}r@{\hspace{4mm}}c@{\hspace{0mm}}c
}
\tabletypesize{\scriptsize}
\tablecaption{2Jy Sample - Continuum features\label{tbl-4}}
\tablewidth{0pt}
\tablehead{
\colhead{Name}{\hspace{1mm}} & \colhead{$z$}{\hspace{0mm}} & \colhead{Optical}{\hspace{0mm}} &
\colhead{Radio}{\hspace{0mm}} & \colhead{$S_{9.7\mu m}$}{\hspace{0mm}} & \colhead{$\pm$}{\hspace{0mm}}& \colhead{$S_{18\mu m}$}{\hspace{0mm}} & \colhead{$\pm$}{\hspace{0mm}}& \colhead{$20/7\mu m$}{\hspace{0mm}}
}

\startdata
0023$-$26	&	\phantom{a}	0.322	&	NLRG	&	CSS	&	-0.33	&	$\pm$	0.15	&	-0.06	&	$\pm$	0.28	&	1.204	\\
0034$-$01	&	\phantom{a}	0.073	&	WLRG	&	FRII	&{0.40	}&	$\pm$	0.06	&	-0.10	&	$\pm$	0.03	&	0.476	\\
0035$-$02	&	\phantom{a}	0.220	&	BLRG	&	(FRII)	&	0.02	&	$\pm$	0.05	&	0.05	&	$\pm$	0.04	&	0.617	\\
0038$+$09	&	\phantom{a}	0.188	&	BLRG	&	FRII	&{0.21	}&	$\pm$	0.02	&	0.04	&	$\pm$	0.02	&	0.598	\\
0039$-$44	&	\phantom{a}	0.346	&	NLRG	&	FRII	&{-0.09	}&	$\pm$	0.02	&	0.02	&	$\pm$	0.02	&	0.722	\\
0043$-$42	&	\phantom{a}	0.116	&	WLRG	&	FRII	&{-0.22	}&	$\pm$	0.02	&	-0.01	&	$\pm$	0.02	&	0.340	\\
0105$-$16	&	\phantom{a}	0.400	&	NLRG	&	FRII	&	0.11	&	$\pm$	0.04	&	-0.12	&	$\pm$	0.06	&	0.610	\\
0117$-$15	&	\phantom{a}	0.565	&	NLRG	&	FRII	&	0.01	&	$\pm$	0.04	&	0.00	&	$\pm$	0.03	&	1.044	\\
0213$-$13	&	\phantom{a}	0.147	&	NLRG	&	FRII	&{-0.37	}&	$\pm$	0.01	&	-0.01	&	$\pm$	0.01	&	0.554	\\
0235$-$19	&	\phantom{a}	0.620	&	BLRG	&	FRII	&{0.24	}&	$\pm$	0.02	&	0.01	&	$\pm$	0.02	&	0.295	\\
0252$-$71	&	\phantom{a}	0.566	&	NLRG	&	CSS	&{0.39	}&	$\pm$	0.07	&	0.04	&	$\pm$	0.05	&	1.004	\\
0347$+$05	&	\phantom{a}	0.339	&	WLRG	&	FRII	&{-0.79	}&	$\pm$	0.06	&	-0.15	&	$\pm$	0.07	&	0.275	\\
0349$-$27	&	\phantom{a}	0.066	&	NLRG	&	FRII	&	0.03	&	$\pm$	0.02	&	-0.02	&	$\pm$	0.04	&	0.438	\\
0404$+$03	&	\phantom{a}	0.089	&	NLRG	&	FRII	&{-1.02	}&	$\pm$	0.08	&	-0.05	&	$\pm$	0.06	&	0.652	\\
0409$-$75	&	\phantom{a}	0.693	&	NLRG	&	FRII	&{-0.33	}&	$\pm$	0.08	&	0.10	&	$\pm$	0.04	&	2.245	\\
0442$-$28	&	\phantom{a}	0.147	&	NLRG	&	FRII	&	-0.04	&	$\pm$	0.02	&	0.02	&	$\pm$	0.02	&	0.337	\\
0620$-$52	&	\phantom{a}	0.051	&	WLRG	&	FRI	&	0.14&	$\pm$	0.06	&	0.02	&	$\pm$	0.04	&	0.239	\\
0625$-$35	&	\phantom{a}	0.055	&	WLRG	&	FRI	&	0.03	&	$\pm$	0.03	&	-0.02	&	$\pm$	0.02	&	0.248	\\
0625$-$53	&	\phantom{a}	0.054	&	WLRG	&	FRII	&	0.20	&	$\pm$	0.13	&	0.18	&	$\pm$	0.14	&	0.103	\\
0806$-$10	&	\phantom{a}	0.110	&	NLRG	&	FRII	&{-0.33	}&	$\pm$	0.02	&	-0.03	&	$\pm$	0.02	&	0.854	\\
0859$-$25	&	\phantom{a}	0.305	&	NLRG	&	FRII	&	-0.09	&	$\pm$	0.04	&	0.04	&	$\pm$	0.05	&	0.533	\\
0915$-$11	&	\phantom{a}	0.054	&	WLRG	&	FRI	&{-0.30	}&	$\pm$	0.04	&	-0.11	&	$\pm$	0.06	&	0.318	\\
0945$+$07	&	\phantom{a}	0.086	&	BLRG	&	FRII	&{0.19}	&	$\pm$	0.04	&	0.00	&	$\pm$	0.03	&	0.389	\\
1136$-$13	&	\phantom{a}	0.554	&	Q	&	FRII	&{0.21	}&	$\pm$	0.02	&	0.00	&	$\pm$	0.02	&	0.431	\\
1151$-$34	&	\phantom{a}	0.258	&	Q	&	CSS	&	0.01	&	$\pm$	0.02	&	0.01	&	$\pm$	0.04	&	0.334	\\
1306$-$09	&	\phantom{a}	0.464	&	NLRG	&	CSS	&	0.05	&	$\pm$	0.05	&	-0.03	&	$\pm$	0.10	&	0.462	\\
1355$-$41	&	\phantom{a}	0.313	&	Q	&	FRII	&{0.20	}&	$\pm$	0.02	&	0.02	&	$\pm$	0.02	&	0.347	\\
1547$-$79	&	\phantom{a}	0.483	&	BLRG	&	FRII	&{0.24	}&	$\pm$	0.04	&	0.01	&	$\pm$	0.06	&	0.671	\\
1559$+$02	&	\phantom{a}	0.104	&	NLRG	&	FRII	&{-0.36	}&	$\pm$	0.02	&	-0.02	&	$\pm$	0.01	&	1.281	\\
1602$+$01	&	\phantom{a}	0.462	&	BLRG	&	FRII	&{0.15	}&	$\pm$	0.03	&	-0.01	&	$\pm$	0.04	&	0.934	\\
1648$+$05	&	\phantom{a}	0.154	&	WLRG	&	FRI	&	--	&			&	--	&			&	--	\\
1733$-$56	&	\phantom{a}	0.098	&	BLRG	&	FRII	&	-0.03	&	$\pm$	0.02	&	0.01	&	$\pm$	0.02	&	0.461	\\
1814$-$63	&	\phantom{a}	0.063	&	NLRG	&	CSS	&{-0.31	}&	$\pm$	0.02	&	-0.05	&	$\pm$	0.04	&	0.472	\\
1839$-$48	&	\phantom{a}	0.112	&	WLRG	&	FRI	&	0.21	&	$\pm$	0.08	&	-0.10	&	$\pm$	0.10	&	0.237	\\
1932$-$46	&	\phantom{a}	0.231	&	BLRG	&	FRII	&	-0.25	&	$\pm$	0.09	&	0.07	&	$\pm$	0.17	&	1.184	\\
1934$-$63	&	\phantom{a}	0.183	&	NLRG	&	GPS	&	0.11	&	$\pm$	0.04	&	-0.02	&	$\pm$	0.03	&	2.122	\\
1938$-$15	&	\phantom{a}	0.452	&	BLRG	&	FRII	&	0.26	&	$\pm$	0.11	&	0.02	&	$\pm$	0.14	&	0.872	\\
1949$+$02	&	\phantom{a}	0.059	&	NLRG	&	FRII	&	0.01	&	$\pm$	0.02	&	-0.05	&	$\pm$	0.02	&	0.606	\\
1954$-$55	&	\phantom{a}	0.060	&	WLRG	&	FRI	&	0.31	&	$\pm$	0.11	&	0.01	&	$\pm$	0.10	&	0.196	\\
2135$-$14	&	\phantom{a}	0.200	&	Q	&	FRII	&	--	&			&	--	&			&	--	\\
2135$-$20	&	\phantom{a}	0.635	&	BLRG	&	CSS	&{-0.51	}&	$\pm$	0.03	&	0.00	&	$\pm$	0.04	&	0.547	\\
2211$-$17	&	\phantom{a}	0.153	&	WLRG	&	FRII	&	--	&			&	--	&			&	--	\\
2221$-$02	&	\phantom{a}	0.057	&	BLRG	&	FRII	&{0.23	}&	$\pm$	0.03	&{0.05}	&	$\pm$	0.01	&	0.297	\\
2250$-$41	&	\phantom{a}	0.310	&	NLRG	&	FRII	&{-0.68	}&	$\pm$	0.01	&	-0.05	&	$\pm$	0.04	&	0.497	\\
2314$+$03	&	\phantom{a}	0.220	&	NLRG	&	FRII	&	0.03	&	$\pm$	0.03	&	0.01	&	$\pm$	0.04	&	2.396	\\
2356$-$61	&	\phantom{a}	0.096	&	NLRG	&	FRII	&{-0.55	}&	$\pm$	0.01	&	-0.03	&	$\pm$	0.01	&	0.895	\\
\enddata

\tablecomments{Table presenting basic parameters of the 2Jy sample as well as spectral continuum feature data. Definitions for column 3 are: NLRG -- narrow-line radio galaxy, BLRG/Q -- broad-line radio galaxy, WLRG -- weak-line radio galaxy, Q -- Quasar. We note that PKS0625-35 is also classified as a BL Lacertae object. Definitions for column 4 are: FRI -- Fanaroff-Riley type 1 object, FRII -- Fanaroff-Riley type 2 object, CSS -- Compact Steep Spectrum object, GPS -- Gigahertz Peaked Spectrum object. Column 5 and 7 present silicate strength defined in Section \ref{sec:silicate}, Equation \ref{eqn:1} for 10 and 18$\mu$m respectively. Silicate detections ($3\sigma$) are marked in bold. Column 9 presents the spectral colour 20/7$\mu$m }  
\end{deluxetable}

\begin{deluxetable}{c@{\hspace{0mm}}c@{\hspace{0mm}}c@{\hspace{0mm}}c@{\hspace{0mm}}r@{\hspace{5mm}}c@{\hspace{1mm}}r@{\hspace{4mm}}c@{\hspace{0mm}}c
}
\tabletypesize{\scriptsize}
\tablecaption{3CRR Sample - Continuum features\label{tbl-5}}
\tablewidth{0pt}
\tablehead{
\colhead{Name}{\hspace{1mm}} & \colhead{$z$}{\hspace{0mm}} & \colhead{Optical}{\hspace{0mm}} &
\colhead{Radio}{\hspace{0mm}} & \colhead{$S_{9.7\mu m}$}{\hspace{0mm}} & \colhead{$\pm$}{\hspace{0mm}}& \colhead{$S_{18\mu m}$}{\hspace{0mm}} & \colhead{$\pm$}{\hspace{0mm}}& \colhead{$20/7\mu m$}{\hspace{0mm}}
}
\startdata
3C33	&	\phantom{a}	0.060	&	NLRG	&	FRII	&{-0.08	}&	$\pm$	0.01	&	0.00	&	$\pm$	0.01	&	0.621	\\
3C35	&	\phantom{a}	0.067	&	NLRG	&	FRII	&	--	&			&	--	&			&	--	\\
3C98	&	\phantom{a}	0.030	&	NLRG	&	FRII	&	-0.18	&	$\pm$	0.07	&	0.02	&	$\pm$	0.05	&	0.772	\\
3C192	&	\phantom{a}	0.060	&	NLRG	&	FRII	&	-0.02	&	$\pm$	0.14	&	-0.07	&	$\pm$	0.23	&	--	\\
3C236	&	\phantom{a}	0.101	&	WLRG	&	FRII	&	0.10	&	$\pm$	0.10	&	0.14	&	$\pm$	0.13	&	0.782	\\
3C277.3	&	\phantom{a}	0.085	&	WLRG	&	FRI/FRII	&	--	&			&	--	&			&	--	\\
3C285	&	\phantom{a}	0.079	&	NLRG	&	FRII	&{-0.54	}&	$\pm$	0.14	&	-0.02	&	$\pm$	0.05	&	1.101	\\
3C293	&	\phantom{a}	0.045	&	WLRG	&	FRI/FRII	&{-1.15	}&	$\pm$	0.03	&	-0.02	&	$\pm$	0.02	&	0.263	\\
3C305	&	\phantom{a}	0.042	&	NLRG	&	CSS/FRII	&{-0.62	}&	$\pm$	0.07	&	-0.07	&	$\pm$	0.15	&	0.828	\\
3C321	&	\phantom{a}	0.096	&	NLRG	&	FRII	&{-0.82	}&	$\pm$	0.04	&	0.01	&	$\pm$	0.01	&	2.659	\\
3C326	&	\phantom{a}	0.090	&	NLRG	&	FRII	&	0.06	&	$\pm$	0.26	&	-0.02	&	$\pm$	0.44	&	0.123	\\
3C382	&	\phantom{a}	0.058	&	BLRG	&	FRII	&{0.23	}&	$\pm$	0.03	&{0.04	}&	$\pm$	0.01	&	0.163	\\
3C388	&	\phantom{a}	0.092	&	WLRG	&	FRII	&	0.23	&	$\pm$	0.18	&	0.13	&	$\pm$	0.20	&	0.192	\\
3C390.3	&	\phantom{a}	0.056	&	BLRG	&	FRII	&{0.13	}&	$\pm$	0.02	&{0.03	}&	$\pm$	0.00	&	0.469	\\
3C452	&	\phantom{a}	0.081	&	NLRG	&	FRII	&{-0.09	}&	$\pm$	0.01	&	0.02	&	$\pm$	0.01	&	0.679	\\
4C73.08	&	\phantom{a}	0.058	&	NLRG	&	FRII	&	--	&			&	--	&			&	--	\\
da240	&	\phantom{a}	0.036	&	WLRG	&	FRII	&	--	&			&	--	&			&	--	\\
\enddata

\tablecomments{Table presenting basic parameters of the 3CRR sample as well as spectral continuum feature data. Definitions are the same as Table \ref{tbl-4}.}  
\end{deluxetable}

\section{Silicate features and the mid-IR continuum}

\subsection{Silicate features}
\label{sec:silicate}

Out of the 56 observed/detected objects in our study, 32 were fitted with a 10$\mu$m silicate feature when modelling with the PAHFIT program. However, not all of these 32 objects show clear evidence for silicates based on visual inspection of the spectra. Therefore, to robustly measure the strengths of the silicate emission/absorption features we adopt the definition of \citet{spoon07}: the silicate strength is the log ratio of the observed flux density at the centre of the silicate feature and the local continuum:

\begin{equation}
  S_{9.7} = \ln \left( \frac{F_{9.7\mu m}[measured]}{F_{9.7\mu m}[continuum]} \right) \label{eqn:1}
\end{equation}

\noindent where $F_{9.7\mu m}[measured]$ and $F_{9.7\mu m}[continuum]$ are the measured flux and
the interpolated continuum flux at the centre of the 9.7$\mu m$ feature respectively; a similar expression applies to the 18$\mu$m feature. To calculate $S_{9.7}$ and $S_{18}$ we first subtracted the PAH features and emission lines from the spectra. Next we measured $F_{9.7/18\mu m}[Measured]$ using a bin width of 0.5$\mu$m. We then calculated ${F_{9.7/18\mu m}[continuum]}$ by fitting a 2nd order polynomial function to the continuum measured from bins of 3$\mu$m wide\footnote{Bin width 6-9$\mu$m and 13-16$\mu$m. } on either side of the silicate feature, and interpolating the fit to estimate the continuum level at the centre of the feature at 9.7/18$\mu$m. Uncertainties on the measured silicate strengths were calculated by combining the uncertainty in
the interpolated continuum measurement and the uncertainty in the flux measured in the spectra at the centre of the silicate feature. The results are presented in Tables \ref{tbl-4} and \ref{tbl-5}, where we have indicated $>$3$\sigma$ silicate detections in bold. Note that, for objects that have strong PAH features which straddle
the 10$\mu$m feature (e.g. PKS2135-20, see below), the silicate strengths are likely to be less accurate, because of the difficulty
of accurately separating the silicate and PAH features.

\begin{figure*}
\epsscale{2.2}
\plotone{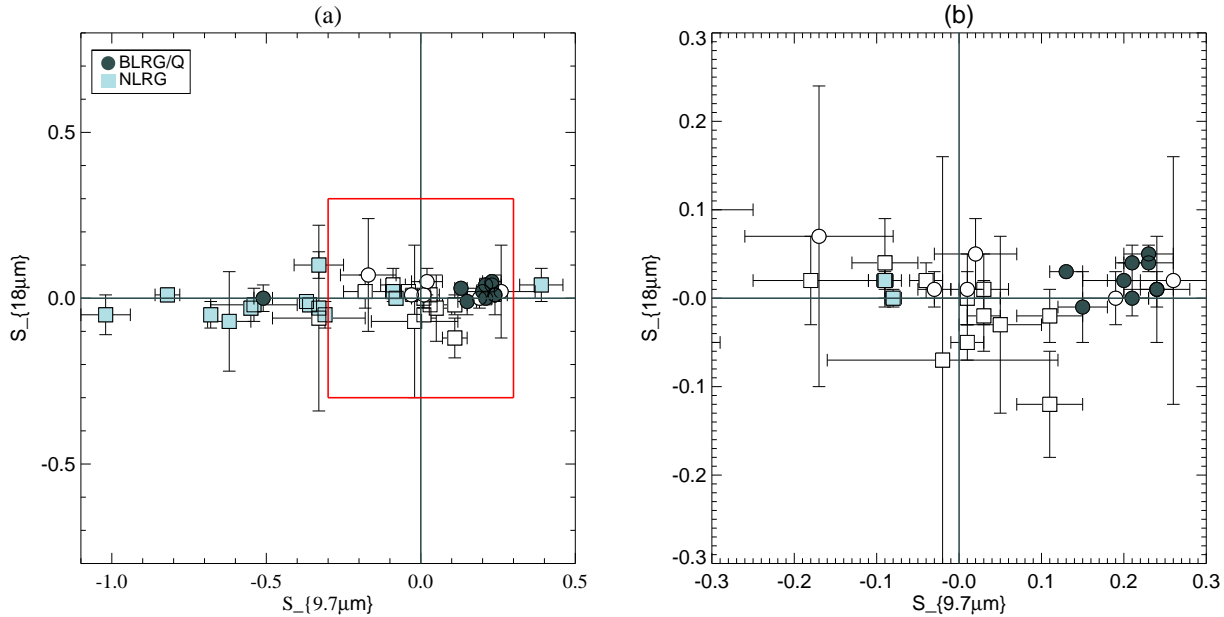}
\caption{: Plots showing $S_{9.7}$ vs $S_{18}$  silicate strength for the BLRG/Q and NLRG objects in the 2Jy and 3CRR samples as defined in Equation \ref{eqn:1}. Filled symbols are 3$\sigma$ detections, open symbols are undetected i.e. below 3$\sigma$. Left plot: All objects with IRS spectra. Right plot: zoom of area marked with red box in left plot. \label{fig:silicate} }
\end{figure*}

Figure \ref{fig:silicate} shows a clear tendency for BLRG/Q to have silicate emission and NLRG to have silicate absorption: 10/11 (91\%) of the objects with significant silicate emission are BLRG/Q, while 18/19 (95\%) with significant silicate absorption are NLRG or WLRG. These  results are consistent with those found in many previous studies of AGN. For example,  \citealp{shi06} -- investigating a mixed sample of Seyferts, quasars and 15 radio galaxies;  \citealp{hao07} -- Seyferts and ULIRGs;  \citealp{baum10} -- Seyferts; \citealp{landt10} -- 12 radio galaxies.  Our results confirm the BLRG/Q to NLRG silicate dichotomy for  a large, complete sample of powerful radio galaxies at intermediate redshifts. 

Secondly, the x- and y-axes of Figure \ref{fig:silicate} (a) are set to an equivalent range in silicate strength. Therefore, it is immediately noticeable that the range of the measured $S_{9.7}$ values is far greater than that of the
measured $S_{18}$ values. This is not surprising because, as described in Section \ref{sec:lines}, when applying the PAHFIT model with silicate absorption we found a much improved fit to the objects using high ratios of 10$\mu$m/18$\mu$m absorption opacity.  

Although in the majority of cases the measurements of the 18$\mu$m silicate feature are consistent with zero within the uncertainties, three objects (3C445, 3C382, 3C390.3) have 18$\mu$m silicate emission detected at a $>$3 sigma level. An emission  feature at 18$\mu$m is clearly visible in the spectra of all these objects. It is also notable that, in general, BLRG/Q with emission at 9.7$\mu$m are fitted better with a PAHFIT model that includes the 18$\mu$m emission feature. 

It is also interesting to consider the detection rate of silicate absorption in the NLRG from the 2Jy and 3CRR samples: 56\%. Given that, based on the unified schemes, we expect circum-nuclear dust structures to obscure the AGN in these objects, it is perhaps surprising that nearly half the NRLG in the samples show no evidence for silicate absorption. In
most cases this lack of strong silicate absorption cannot be attributed to low S/N. Rather, this
result may indicate dilution of the mid-IR continuum  by non-torus sources of emission.
Possibilities include emission from dust in the NLR (e.g. \citealp{radomski02}) that is out of the plane of the torus, and/or non-thermal synchrotron emission, but the lack of detections may also reflect the true diversity in the mid-IR emitting dust structures and/or the range in the angle of such structures to the line of sight.

Looking at Figure \ref{fig:silicate} (b) it is clear that the scatter of silicate emission values for the BLRG/Q with significant silicate emission detections\footnote{Note the BLRG/Q without silicate detection have either strong PAH features (PKS1151-34, PKS1733-56, PKS2135-20) that can mask the emission feature, noisy spectra (PKS1938-15), or calibration uncertainties (PKS0035-02).} is much smaller than that of the silicate absorption values:  whereas the measured silicate emission strengths ($S_{9.7}$) for the BLRG/Q range
from 0.13 to 0.24 (i.e. a spread of 0.11),  the silicate absorption strengths measured in 
the NLRG range from  -0.09 to -1.02 (a spread of approximately 1.0). It is possible
that the range of silicate absorption depths of the NLRG reflects varying lines of sight through
the central obscuring dust structures, or varying degrees of dilution by non-torus continuum components. However, although we expect the range of emission features to be less than that of the absorption, the narrow range of $S_{9.7}$ emission measured for the majority of BLRG/Q suggests remarkable uniformity in the mid-IR emission mechanism in such objects. 

There are two notable outliers in silicate strength compared to the general trends presented above. PKS2135-20 is a BLRG with apparently strong silicate absorption. Although silicate absorption cannot be
entirely ruled out for this object, the strength of the PAH features detected in its IRS spectra makes the contribution of silicate absorption difficult to accurately quantify, because the PAH emission dominates the wings of the putative silicate absorption feature. In contrast, the compact steep spectrum object PKS0252-71 has the highest $S_{9.7}$ {\it emission} strength measured for any object in the two samples, despite the fact that it is a NLRG; it is not clear why a NLRG should have stronger silicate emission than seen for BLRG/Q objects.
 
We further note that, for all but one (PKS0235-19) of the objects with significant silicate emission,  we find that the peak of the 9.7$\mu$m silicate emission feature is redshifted. The wavelength range of the redshifted peak is 10.0-10.6$\mu$m, but the shift is not seen in the 18$\mu$m  feature (where detected). Red shifts in the silicate emission feature have been measured in several other Spitzer studies of AGN (e.g. \citealp{shi06}; \citealp{landt10}; \citealp{leipski10}), but there is currently no general agreement about the cause of the shift, which has been variously attributed to obscuration effects, variations in the dust grain size or dust chemistry, radiative transfer effects, or a shift in the blackbody spectrum due to temperature \citep{leipski10}. It is also important to note that $S_{9.7}$ emission strength may be slightly underestimated in Tables \ref{tbl-4} and \ref{tbl-5} because we did not allow for any shift in the peak of this feature in our measurements, and therefore the silicate emission strength is not measured at the peak of emission.

\subsection{Spectral shape}
\label{sec:shape}

\begin{figure*}[t]
\epsscale{2.2}
\plotone{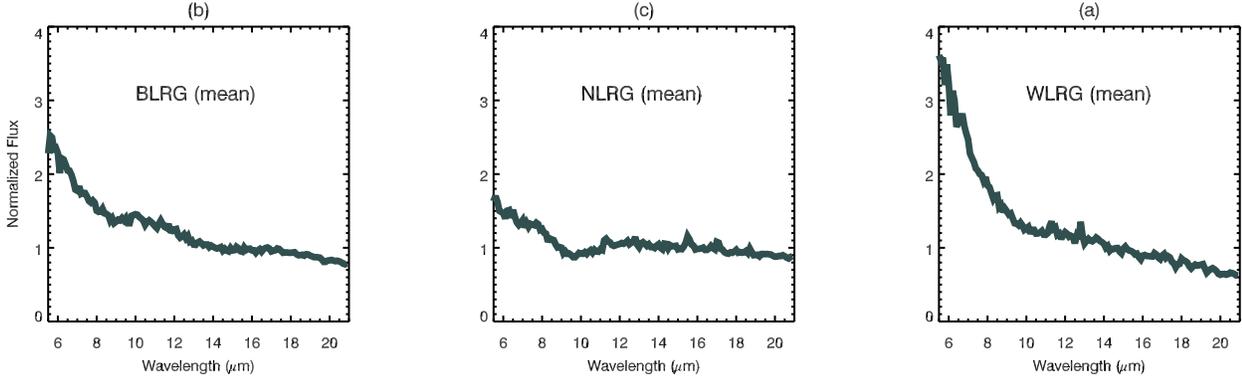}
\caption{: Mean spectra divided into optical class for the combined 2Jy and 3CRR samples. All three spectra have been normalised so that they have the same relative flux at 14 microns.\label{fig:mean} }
\end{figure*}

It is interesting to consider whether there are any correlations between
the overall mid-IR continuum shapes and the strengths of the silicate features. For example, in
the simplest case of a screen of absorbing dust between the mid-IR continuum source and
the observer, it would be expected that, due to extinction effects, the continuum would become redder as
the silicate absorption depth increased. 

In Figure \ref{fig:mean} we plot mean spectra divided into optical classifications. After subtracting emission line and PAH features the mean spectra were created by re-sampling the  data in wavelength bins and normalising all the spectra to have the same flux level at 20$\mu$m.  The results clearly show the silicate feature is in emission for BLRG/Q, and in absorption for NLRG, but the appearance of these features in the mean spectra is likely to be
a consequence of objects with strong silicate features dominating the mean, and does not reflect the fact that silicate features are not detected in a significant fraction ($\sim$40\%) of the spectra of the individual objects. Looking at the overall spectral shape, the mean spectra of the three optical classes show the largest contrast towards the blue end of the spectrum, in particular 
the mean spectrum of the WLRG. This can be attributed to the stellar contribution from the host galaxies, which is likely to dominate the short wavelength continuum for these nearby objects, which have weaker AGN. However, the mid-IR continuum shape of the BLRG/Q is also notably bluer than that of the NLRG, as expected if the latter objects are affected by a higher level of dust extinction. 

\begin{figure}[h]
\epsscale{1}
\plotone{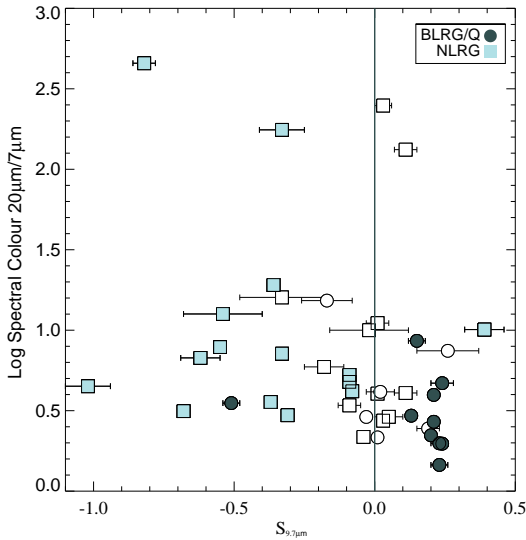}
\caption{: Plot of $S_{9.7}$ versus IRS spectral colour 20/7$\mu$m . Filled squares are narrow-line radio galaxies; filled circles are broad-line radio galaxies. Unfilled objects are undetected in $S_{9.7}$ i.e. below 3$\sigma$.  \label{fig:SilicateColour} }
\end{figure}

 
To further investigate the mid-IR continuum shapes,  we have measured the 20/7$\mu$m continuum
colours from the spectra of individual objects, following
the subtraction of all emission lines, PAH and silicate features using the PAHFIT model. Figure \ref{fig:SilicateColour} 
shows the 20/7$\mu$m colour plotted against the 9.7$\mu$m silicate strength ($S_{9.7}$). Although some of the BLRG/Q objects
with the strongest 9.7$\mu$m emission features also have the bluest 20/7$\mu$m colours, and some of the NLRG with
the strongest 9.7$\mu$m absorption features have the reddest colours, overall, the plot shows a large scatter {(see also \citealp{ogle06})}. This lack of
a clear-cut correlation between the silicate feature strength and the mid-IR colour argues against the naive expectations
of a simple dust screen extinction model.

\begin{figure}[h]
\epsscale{1}
\plotone{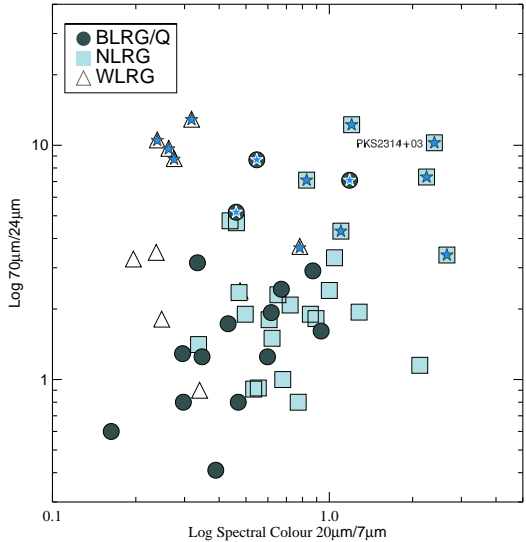}
\caption{: Plot of the IRS spectral colour 7$\mu$m /20$\mu$m versus mid- to far-infrared colour (70/24$\mu$m). Filled squares are narrow-line radio galaxies; filled circles are broad-line radio galaxies; open triangles are weak-line radio galaxies. In addition, objects with evidence for recent star formation activity are indicated with a blue star. \label{fig:colour} }
\end{figure}

One possible reason for the lack of a correlation in Figure \ref{fig:SilicateColour} is that non-torus emission
components contribute to the mid-IR continuum. As well as the possibility of NLR emission components discussed in Section \ref{sec:silicate}, starburst heated dust may also contribute to the longer wavelength mid-IR emission. In order to investigate the significance of the latter, in Figure \ref{fig:colour} we plot 20/7$\mu$m  versus mid- to far-infrared colour (70/24$\mu$m; measured from Spitzer/MIPS photometric data - \citealp{dicken10}).  Once again we have labelled the objects according to optical class and, in addition, we have indicated objects that show good, independent, evidence for recent star formation activity (RSFA) at optical and mid-IR wavelengths, as defined in \citet{dicken12}. Figure \ref{fig:colour} shows that 70/24$\mu$m colour ratio is an excellent indicator of RSFA (see also \citealp{dicken09,dicken12}). However, this plot does not reveal any clear relationship between the shape of the mid-IR spectrum and the presence of RSFA for the sample as a whole, apart from the fact that 3/4 of the objects with extremely red 20/7$\mu$m colours also
show evidence for RSFA and have red, starburst-like 70/24$\mu$m MFIR colours. One of the
latter objects, the ULIRG PKS2314+03 (identified in Figure \ref{fig:colour}) has unusually red mid-IR and MFIR continuum colours, and this result is paralleled in Figure \ref{fig:O4plot}(b) where PKS2314+03 stands out as the only NLRG lying above the correlation between $L_{[\rm{OIV}]}$ and $L_{24\mu m}$. Therefore, it is probable that the RSFA in PKS2314+03 is making a significant contribution to its 24$\mu$m emission. 
Overall, while the presence of RSFA is likely to increase the scatter in the mid-IR colours, it is unlikely
to provide the full explanation for the lack of correlation in Figure 8, since the overwhelming
majority ($>$65\%) of objects in our samples show no evidence for significant RSFA \citep{dicken12}.

\section{Discussion}
\label{sec:disc}

\subsection{The origin of the mid-IR emission lines}
\label{sec:star}

The higher ionisation  fine structure lines investigated so far (e.g.  [OIV]: $E_{ion} = $54.9eV; [NeIII]: $E_{ion} = $41.0eV) are not only thought to originate predominantly from the narrow-line regions of AGN (\citealp{melendez08}; \citealp{rigby09}; \citealp{diamond09}), but are also fainter in pure starburst galaxies, where they can only arise from shocks associated with very hot stars \citep{lutz98}. In contrast, the lower ionisation fine structure lines ( $E_{ion}\leq$30eV) can be readily produced by stellar photoionisation in HII regions rather than by AGN. Indeed, previous studies used the relative strength of the high and low ionisation lines  to gauge the relative dominance of starburst and AGN in Seyfert and ULIRG objects (e.g. \citealp{sturm02}; \citealp{veilleux09a}). Because optical line ratios become unreliable for heavily obscured objects, mid-IR diagnostics are an important alternative for determining the dominant ionisation mechanism. 

Out of the 56 observed/detected objects in the 2Jy and 3CRR samples we detect [NeII] $\lambda$12.81$\mu$m ($E_{ion}$=21.6eV) in 39 objects (70\%). This line has the second lowest ionisation energy amongst the mid-IR fine-structure lines detected and presented in this work, the lowest being the weaker [ArIII] $\lambda$6.98$\mu$m line ($E_{ion}$=15.8eV), which suffers from a much lower detection rate (15/56, 27\%). 

\begin{figure}
\epsscale{1}
\plotone{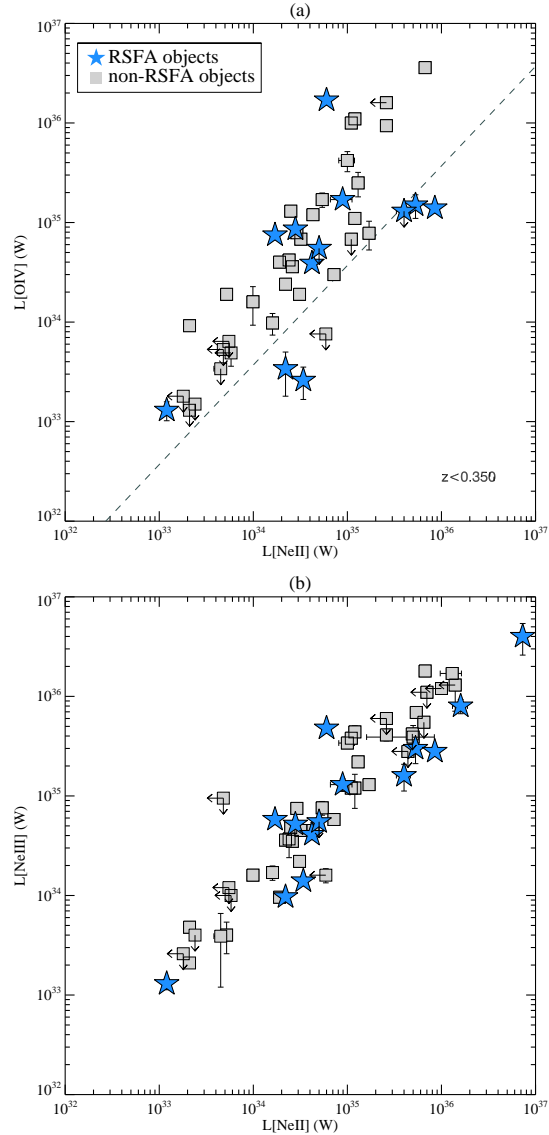}
\caption{Plots of $L_{[\rm{NeII}]}$ versus $L_{[\rm{OIV}]}$ and $L_{[\rm{NeIII}]}$ with object identified with recent star formation activity (RSFA) marked as stars. These plots includes strong-line and weak-line objects. The line plotted in (a) represents the critical line ratio ($[NeII]/[OIV] = 2.7$) that marks the boundary between
AGN or starburst origin to the infrared emission lines. Objects below this line are expected to have a star formation origin.  \label{fig:SBdiag} }
\end{figure}

To further investigate the origin of the mid-IR lines in Figure \ref{fig:SBdiag} we plot $L_{[\rm{NeII}]}$ versus $L_{[\rm{OIV}]}$ and $L_{[\rm{NeIII}]}$, including both the strong-line and weak-line radio galaxies, and again identifying objects with evidence for RSFA at optical and mid-IR wavelengths.  

Studying Figure \ref{fig:SBdiag}(a) and (b), both $L_{[\rm{OIV}]}$ and $L_{[\rm{NeIII}]}$ appear to correlate well with $L_{[\rm{NeII}]}$. Given the evidence presented in this paper for an AGN origin of both the [OIV] and [NeIII] emission lines, it follows that the [NeII] emission line is also likely to have an AGN origin in most objects. However, mid-IR lines such as [NeII] have been shown to be produced predominantly by star formation in studies of Seyfert galaxies (e.g. \citealp{veilleux09a}), therefore we should consider the possibility that star formation contributes to this emission line. In fact, it is notable that $L_{[\rm{NeII}]}$ is detected in all but one\footnote{3C236, which has the lowest equivalent width detection of the star formation tracing PAH and second warmest colour ratio (70/24$\mu$m=3.4) of all the RSFA objects} of the RSFA identified objects. 

Considering Figure \ref{fig:SBdiag}(a), the three RSFA objects with the highest $L_{[\rm{NeII}]}$ luminosities, lying in a small group below the apparent correlation -- PKS0023-46, PKS0347+05, PKS2314+03 --  all have red mid-infrared spectral continuum colours and the \emph{coolest} 70/24$\mu$m colours as well as strong PAH features. Moreover, the two RSFA identified objects, lying below the apparent correlation towards the centre of Figure \ref{fig:SBdiag}(a) are the weak-line objects PKS0915-11 (Hydra A) and 3C293. These two objects have similar IRS spectra, showing weak continua at longer mid-IR wavelengths, very strong PAH features, and high equivalent width [NeII] emission. They also show high equivalent width detections of the low ionisation emission line [ArIII] $\lambda$6.98$\mu$m. 

Based on a study of well-known, nearby, Seyfert objects \citet{sturm02} proposed that objects with values of the $[NeII]/[OIV]$ ratio lower than or equal to 2.7 are likely to have a dominant AGN origin; whereas those with higher ratios are
likely to have a significant contribution from starburst heating. We have plotted the line that represents the 
critical ratio in Figure \ref{fig:SBdiag}(a). Considering the $[NeII]/[OIV]$ ratio for the combined 2Jy and 3CRR sample\footnote{Two RSFA objects (PKS2135-20; PKS0409-75) could not be included in this analysis due to their [OIV] emission line being redshifted out of the rest wavelength range of the spectra.}, the majority of objects have $[NeII]/[OIV]$  ratios below 2.7, and the mean is $[NeII]/[OIV] = 0.70\pm0.63$ for the non-RSFA sample. This supports an AGN origin for the mid-IR emission lines in the majority of the radio galaxies in the combined 
sample. However, the five RSFA objects lying below the correlation in Figure \ref{fig:SBdiag}(a), mentioned above, all have $[NeII]/[OIV] > 2.7$. Given the independent evidence for these five objects to have the strongest RSFA in our samples, the
$[NeII]/[OIV]$ ratio demonstrates to be an effective diagnostic to gauge dominant star formation activity in powerful radio galaxies. 
 
Returning to Figure \ref{fig:SBdiag}(b), which plots $L_{[\rm{NeII}]}$ against $L_{[\rm{NeIII}]}$, it is interesting that, although many of the RSFA objects lie at the bottom edge of the apparent correlation, there are no objects that show high ratios of [NeII]/[NeIII] and fall well below the correlation, seen in Figure \ref{fig:SBdiag}(a) for the [NeII]/[OIV]  ratio; it is also clear that the degree of scatter is lower in the $L_{[\rm{NeII}]}$ versus $L_{[\rm{NeIII}]}$ plot. These 
results can be explained if star formation makes a significant contribution to the [NeIII], as well as the [NeII] emission, thus diluting the [NeII]/[NeIII] ratio as a useful AGN or star formation diagnostic. This explanation is also consistent with the results from Section \ref{sec:opt} which show that $L_{[\rm{NeIII}]}$ is not as reliable a tracer of AGN power as $L_{[\rm{OIV}]}$. 

In summary, a maximum of 35\% of the powerful radio galaxies show evidence for RSFA, and even fewer (8\%) show evidence that the star formation activity contributes significantly to the production of the mid-IR emission lines. It is therefore likely that, for the majority of powerful radio galaxies, both the high and low ionisation mid-IR lines are dominated by AGN photoionisation. On the other hand, it is clearly important to be cautious about using the [NeIII], and particularly the [NeII], fine structure lines as indicators of AGN power in objects that show independent evidence for RSFA.

\subsection{Unification}
\label{sec:uni}

The idea of an orientation-based unified scheme for the optical classes of radio sources was proposed over two decades ago (e.g. \citealp{barthel89}) with the BLRG/Q proposed to represent a population with the  torus axis oriented close to the line of sight, allowing a full view of the broad line and continuum emitting regions, and the NLRG to represent a heavily obscured population in which the axis of the torus is oriented at a large angle to the line of sight. Optical polarisation studies of radio-loud AGN have provided the evidence that NLRG and BLRG/Q indeed represent the same population viewed from different directions {(e.g. \citealp{antonucci84,cohen99})}. Mid-infrared spectra are ideal for further testing such unified schemes because the emission is less likely to suffer from obscuration compared to optical wavelength studies. {Following on from previous investigations of unification using IRS spectra (e.g. \citealp{ogle06,shi06,leipski09}), with the new 2Jy sample IRS data, and the extensive archive of complementary multi-wavelength data for this sample, we are able to further test the orientation-based unified schemes.  }

We have presented evidence for obscuration as predicted by the unified schemes, albeit mild and with a wide dispersion. Median ratio values for the two optical classes of radio sources are [OIII]/[OIV] = 1.6(BLRG/Q) and 0.7(NLRG); similar results have been found for type 1 and type 2 Seyfert galaxies (e.g. \citealp{baum10}; \citealp{lamassa10}) and for radio galaxies (e.g. \citealp{jackson97}; \citealp{haas05}). For Seyfert type 1 and 2 galaxies \citet{lamassa10} also investigated [OIII]/[OIV] and found a difference of up to a factor of two, on average, between the [OIII] luminosities of type 1 and type 2 objects, in good agreement with our results. In a previous study of radio galaxies, \citet{haas05} found evidence for a larger average difference, with the [OIII]/[OIV] ratio a factor of approximately ten lower for the NLRG compared with the BLRG/Q in their sample. In contrast, although there is large dispersion, our data show a median ratio between BLRG/Q and NLRG [OIII]/[OIV] values of only 2.3. However, the sample of \citet{haas05} was statistically much smaller than that used in this study, comprising of only seven NLRG and seven quasars, matched in radio luminosity. 

\citet{cleary07} also investigated the mid-IR properties of 3CR radio galaxies and quasars at intermediate redshifts and present evidence for a difference in rest frame 15$\mu$m luminosity. They found a difference between the mean $R_{dr} =  \nu L_{\nu}(IR)/\nu L_{\nu}(178 MHz)$ for narrow-line radio galaxies compared to quasars of a factor of 2, when corrected for non-thermal contamination. They attribute this to obscuration of the mid-IR continuum emission. Despite the fact that there is an uncertainty in the $R_{dr}$ value due to the variation in $L_{(178 MHz)}$ as described in Section \ref{sec:AGNpower}, the implied level of attenuation of the mid-IR continuum is similar to that deduced in this study on the basis of the 24/[OIV] ratio. 

The evidence for obscuration and dependence on orientation of the optical classes of radio galaxies is also supported by the clear divide between detections of BLRG/Q with silicate emission and NLRG with silicate absorption. This result is well established for the Seyfert AGN population, and is also predicted by clumpy torus models \citep{nenkova02,nenkova08}, where silicate emission in BLRG/Q is associated with hot dust close to the active nucleus and the silicate absorption in NLRG is due to circum-nuclear dust that absorbs the AGN continuum along certain lines of sight. 

\begin{figure}[h]
\epsscale{1}
\plotone{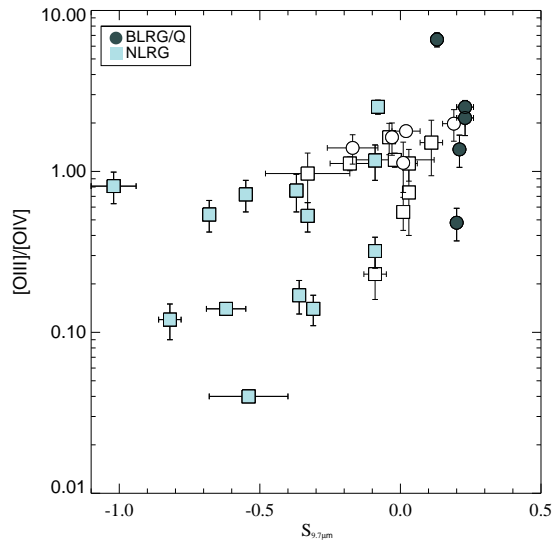}
\caption{: Plot of the $S_{9.7}$ versus [OIII]/[OIV]. Filled squares are narrow-line radio galaxies; filled circles are broad-line radio galaxies. Unfilled objects are undetected in $S_{9.7}$ i.e. below 3$\sigma$.  \label{fig:SilicateRatio} }
\end{figure}

In Figure \ref{fig:SilicateRatio} we investigate the possibility that objects showing evidence for attenuation of the [OIII] optical emission line also show silicate absorption. We emphasise that a correlation is not necessarily expected because of the potential contamination of the mid-IR continuum by various non-torus continuum components, and significant variations in the structure of the circum-nuclear dust structures from object to object. Also, the [OIII]/[OIV] ratio will depend on the ionisation state of the gas, and any differences in density, radial gas distribution and metallicity. However, despite the scatter, a weak trend towards stronger silicate absorption for low ratios of [OIII]/[OIV] can be identified. This add weight to the hypothesis that obscuration is the main cause of the difference seen between the [OIII] luminosities of NLRG and BLRG/Q. 

A prediction of clumpy torus models is that the silicate absorption should be relatively shallow, with $S_{9.7} > -1.0$ \citep{nenkova08}. This is entirely consistent with our results. However, as well as the relatively shallow absorption features, the detection rate of silicate absorption features in only 56\% for the 9.7$\mu$m, and zero for the 18$\mu$m feature. The lack of detections of silicate absorption features in $>$40\% of the NLRG is all the more surprising given that the kpc-scale disks may contribute to the measured silicate absorption over-and-above that produced by a compact torus structure. \citet{gallimore10} also found evidence that strong silicate features are not ubiquitous in AGN: for 35\% of the Seyfert galaxies in their sample the 9.7$\mu$m feature is not detected at greater than $3\sigma$. They found that objects without silicate detections tend to belong to particular sub-samples of the Seyfert population, including Seyfert 2 objects without evidence for hidden broad-line regions, intermediate Seyfert types (S1.8-1.9), and LINERs. We find no trend in the 2Jy and 3CRR samples between silicate detection or strength with either the object classifications or their properties. 

Models predict that clumpy tori with low average numbers of clouds along the line of sight ($N_0 \le 2$) never produce an absorption feature, regardless of the orientation, and  that the silicate features commonly detected in Seyfert spectra are reproduced by models that include $N_0 \ge 5$ clouds (\citealp{nenkova08}; \citealp{alonso-herrero11}). Overall, if we assume that the mid-IR continuum is solely due to the torus, the silicate emission (BLRG/Q) and absorption (NLRG) strengths measured in our sample are consistent with clumpy models that have relatively few clouds along the line of sight $2 < N_0 < 5$. However, the large number of non-detections of silicate absorption features is also consistent with the idea that much of the mid-IR continuum does not suffer extinction by the torus because it is emitted by structures on larger scales such as the NLR, circum-nuclear starbursts or synchrotron-emitting jets.

\subsection{Weak-line radio galaxies}
\label{sec:weak}

\begin{figure*}
\epsscale{1.9}
\plotone{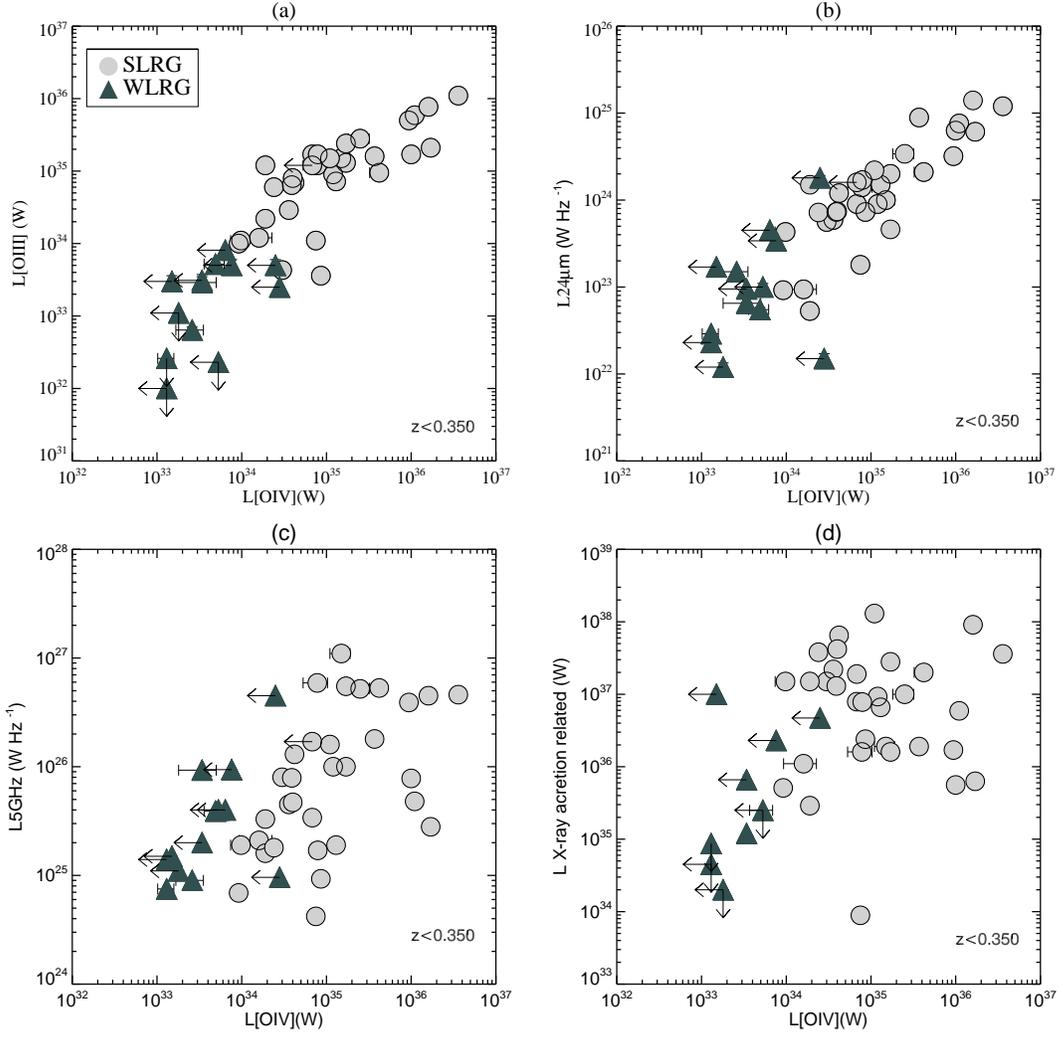}
\caption{Plot showing $L_{[\rm{OIV}]}\lambda$25.89$\mu$m versus $L_{[\rm{OIII}]}\lambda5007$, $L_{24\mu m}$, $L_{5GHz}$  and $L_{[\rm{NeIII}]}\lambda$15.56$\mu$m with objects identified by Optical classes including weak-line radio galaxies (WLRG). The objects plotted are limited to objects with $z$$<$0.350 due to the [OIV] line redshifted out of the redshift range of Spitzer/IRS. SLRG --- strong-line radio galaxies/quasars.  \label{fig:wlrg} }
\end{figure*}

In terms of the classification of radio-loud AGN, there has recently been much debate about the nature of weak line radio galaxies (WLRG), and how they relate to strong line (NLRG and BLRG/Q) objects of similar radio power (e.g. \citealp{hardcastle07}; \citealp{buttiglione10}; \citealp{smolcic11}). One explanation for the relatively low emission line luminosities of WLRG is that their AGN have intrinsically low radiative powers and therefore do not produce luminous, high ionisation emission line regions by illuminating the ISM in the host galaxies, as is assumed to happen in strong-line radio galaxies (SLRG). Alternatively, it has been
proposed that the relatively low emission line luminosities of the WLRG are due to an unusually
high level of dust extinction affecting the NLR. 

\citet{dicken09,dicken10} and \citet{hardcastle09} showed that the WLRG not only have low [OIII] emission line luminosities,
but also low 24$\mu$m continuum luminosities. This already suggests that WLRG are unlikely to harbour
luminous quasar-like nuclei that are obscured by dust at optical wavelengths, {as also supported by the study of \citet{ogle06}}. However, as we have already seen in this study, the 24$\mu$m emission can itself suffer from significant extinction by the circum-nuclear dust; if such obscuration is particularly extreme in the WLRG compared with their SLRG counterparts, this could potentially explain the lower 24$\mu$m luminosities of the WLRG.

We can re-consider this question using the new mid-IR emission line data. Figure \ref{fig:wlrg} again shows the correlations between the AGN power indicators, but now including the WLRGs plotted alongside the SLRG. Inspecting this figure, it is clear that the WLRG fall at the low luminosity end of all the correlations, and make up the majority of the objects with upper limits in [OIV]. The low detection rate of the WLRG using this mid-IR AGN power proxy (only 16\%) reinforces our previous conclusion that WLRG have intrinsically low power AGN; since the [OIV] emission suffers relatively little from the effects of dust extinction compared to the optical [OIII] line, these results strongly reinforce the idea that the WLRG have intrinsically low radiative powers. 

\section{Conclusions}

We have investigated deep Spitzer/IRS spectra for complete samples of 46 2Jy radio galaxies (0.05$<z<$0.7) and 17 3CRR FRII radio galaxies ($z<$0.1). Our main conclusions are as follows.

\begin{itemize}

\item{The luminosity of the mid-IR [OIV] $\lambda$25.89$\mu$m emission line is at least as reliable a tracer of AGN power as the optical [OIII] $\lambda$5007 emission line and 24$\mu$m mid-IR continuum luminosities, but has the added notable advantage that
it suffers less from dust extinction. However, total radio power (at 5GHz) and X-ray luminosity do not perform as well when compared with mid-IR [OIV] $\lambda$25.89$\mu$m and optical [OIII] $\lambda$5007 emission line luminosity as proxies of AGN power for powerful radio galaxies.  }

\item{Based on comparisons between the BLRG/Q and NLRG objects in correlation plots that include the mid-IR
fine-structure lines ([OIV] $\lambda$25.89$\mu$m and [NeIII] $\lambda$15.56$\mu$m), the optical [OIII] $\lambda$5007 line emission and the mid-IR 24$\mu$m continuum emission both show evidence for a mild degree of attenuation of a  factor $\approx$2. We attribute this to obscuration by circum-nuclear dust structures. However, the degree of attenuation is considerably lower than claimed in some previous studies of powerful radio galaxies. }

\item{10$\mu$m silicate features are detected in approximately 60\% of the objects and therefore do not appear to be ubiquitous in powerful radio galaxies. Silicate emission is predominantly detected in BLRG/Q, with a detection rate of 59\%, whereas silicate absorption is only detected in NLRG with a detection rate of 56\%. This dependence on optical classification is consistent with many previous investigation of AGN. We find no obvious relationship between silicate detection or non-detection with any other observable properties from the two samples.  }

\item{From an analysis of the ratios of mid-IR emission lines we conclude that these lines are produced
by AGN photoionisation in the majority of powerful radio galaxies. However, in the small minority of objects with evidence for the strong star formation activity, stellar photoionisation can make a significant contribution to the lower ionisation mid-IR emission lines (e.g. [NeII] $\lambda$12.81$\mu$m and [NeIII] $\lambda$15.56$\mu$m). }

\item{Using the isotropic [OIV] emission line as an AGN power indicator, we provide further evidence against the idea that weak-line radio harbour intrinsically powerful, but heavily extinguished AGN.  }

\end{itemize}

{\it Facilities:} \facility{Spitzer (IRS)}

\acknowledgments ({We thank the anonymous referee for their comments which aided this investigation and helped put the work in context.})We would like to thank Jack Gallimore for assistance with the modification of PAHFIT. This work is based [in part] on observations made
with the Spitzer Space Telescope, which is operated by the Jet
Propulsion Laboratory, California Institute of Technology under a
contract with NASA. Based on observations made with ESO Telescopes at the Paranal Observatory.
D. D. acknowledges support from NASA grant based on observations from 
Spitzer program 50588 and the NASA ROSES ADAP program. CRA acknowledges financial support from the Instituto de Astrof\' isica de Canarias and the Spanish Ministry of Science and Innovation (MICINN) through project PN AYA2010-21887-C04.04 (Estallidos). M.B.N.K. was supported by the Peter and Patricia Gruber Foundation through the PPGF fellowship, by the Peking University One Hundred Talent Fund (985), and by the National Natural Science Foundation of China (grants 11010237, 11050110414, 11173004). This publication was made possible through the support of a grant from the John Templeton Foundation and National Astronomical Observatories of Chinese Academy of Sciences. The opinions expressed in this publication are those of the author(s) do not necessarily reflect the views of the John Templeton Foundation or National Astronomical Observatories of Chinese Academy of Sciences. The funds from John Templeton Foundation were awarded in a grant to The University of Chicago which also managed the program in conjunction with National Astronomical Observatories, Chinese Academy of Sciences.

\bibliographystyle{apj.bst} 
\bibliography{bib_list}

\appendix

\begin{figure*}[t]
\epsscale{1}
\plottwo{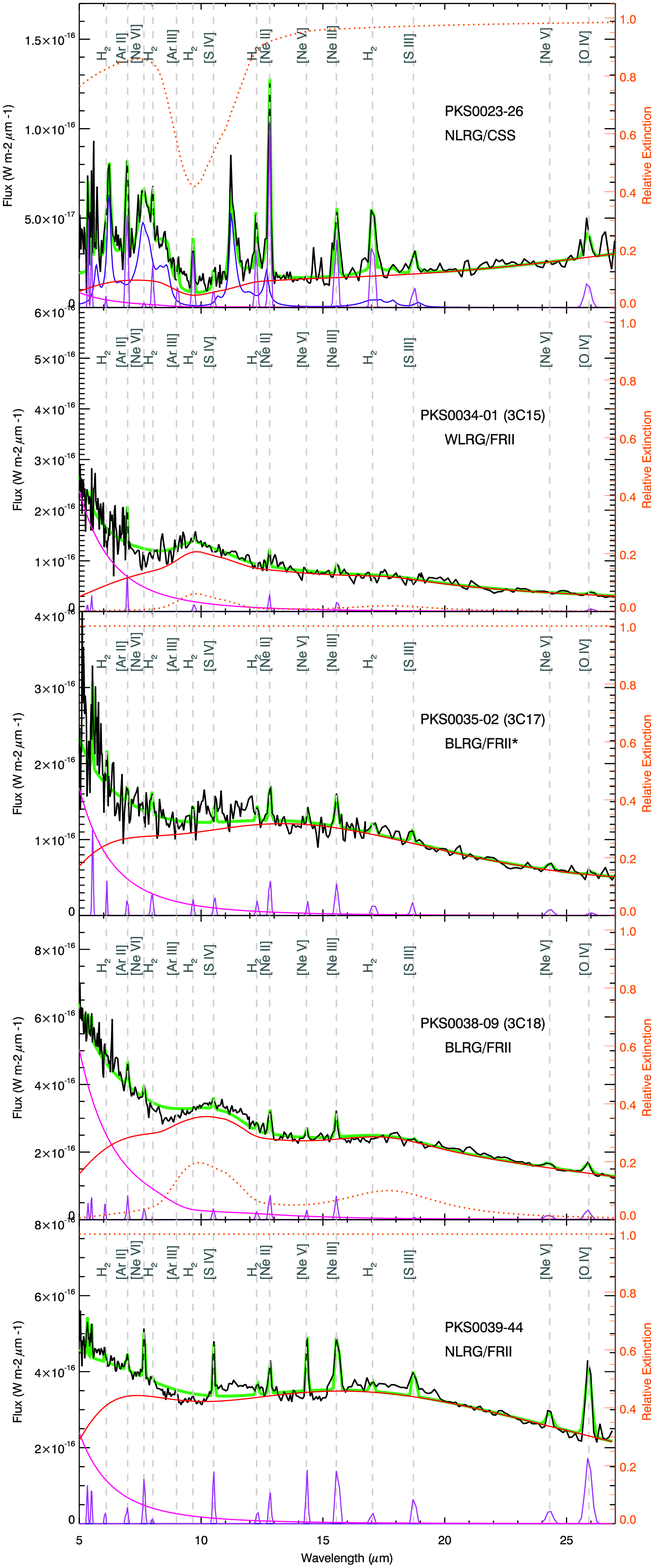}{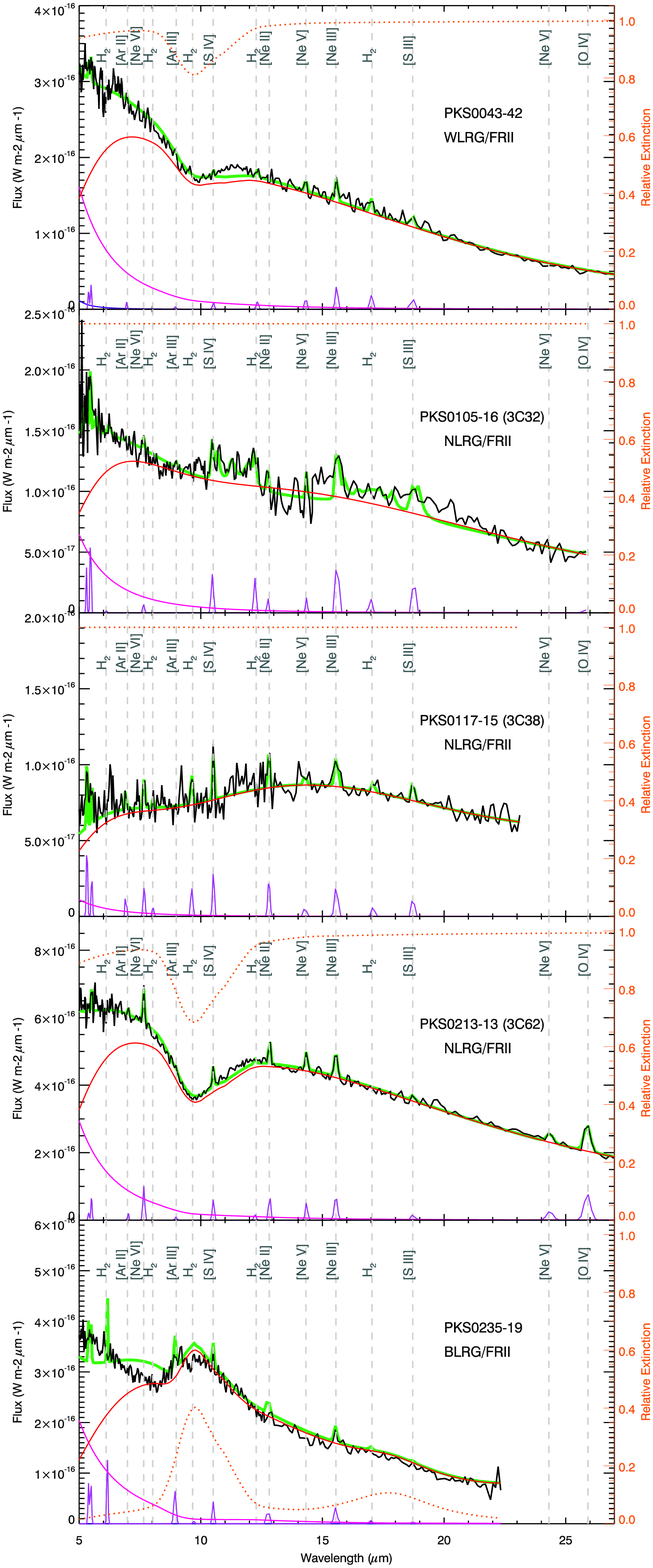}
\caption{Spitzer/IRS spectra for the 2Jy sample. {Note that all spectra were previously published in \citet{dicken12} and 4 spectra were first published in \citet{ogle06,leipski09}}. Common fine structure and H$_2$ emission lines are indicated by vertical grey dashed lines. The dotted orange line shows the extinction ($e^{- \tau \nu}$ = 1 if no extinction). Prominent PAH features are indicated by the blue line. Note that data for PKS0034-01 and PKS0035-02 potentially suffer from enhanced flux calibration uncertainties at short wavelengths, due to saturation of the peak-up detector (see \citet{dicken12} for details). The SL continuum agreement between nod positions is not good for PKS0039-44 between the ranges 10-14$\mu$m likely leading to the strange shape of the spectrum. {3C15 also published in \citet{leipski09}.} \label{spec1} }
\end{figure*}

\begin{figure*}[t]
\epsscale{1}
\plottwo{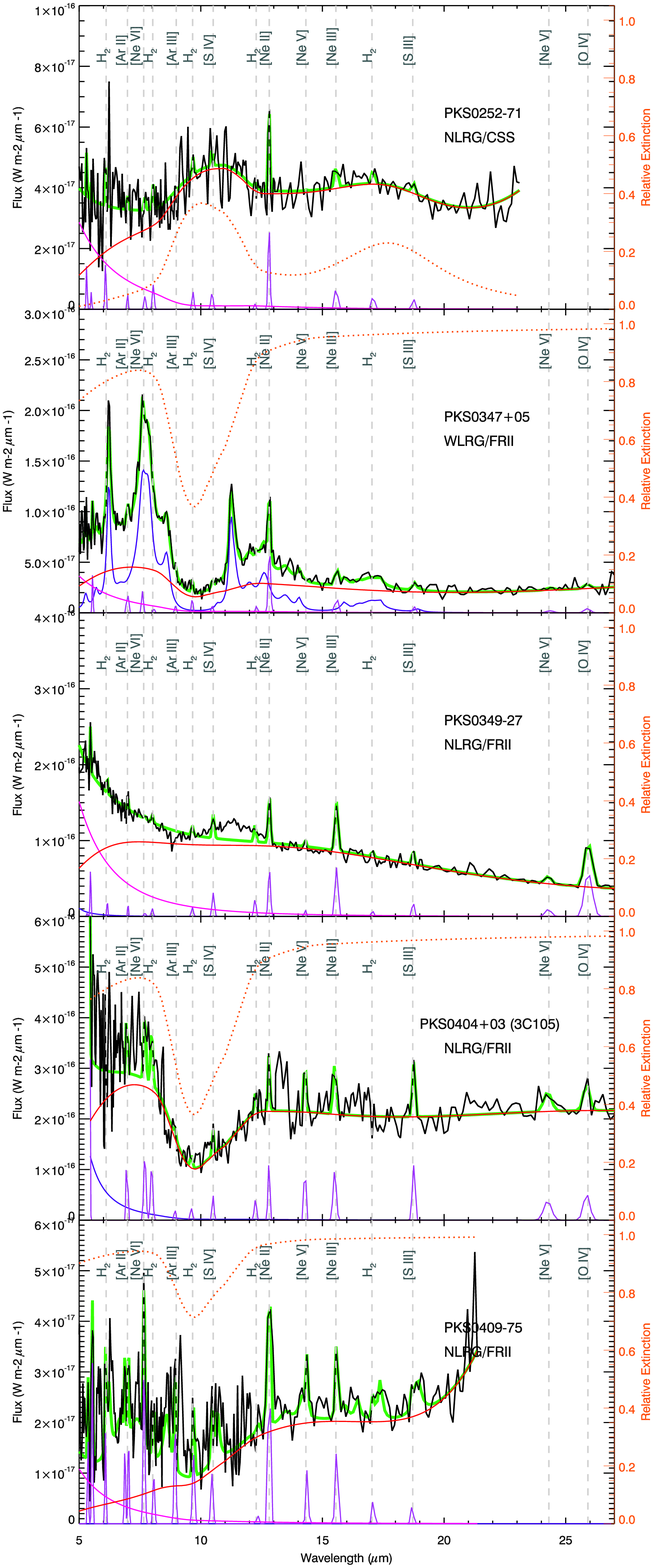}{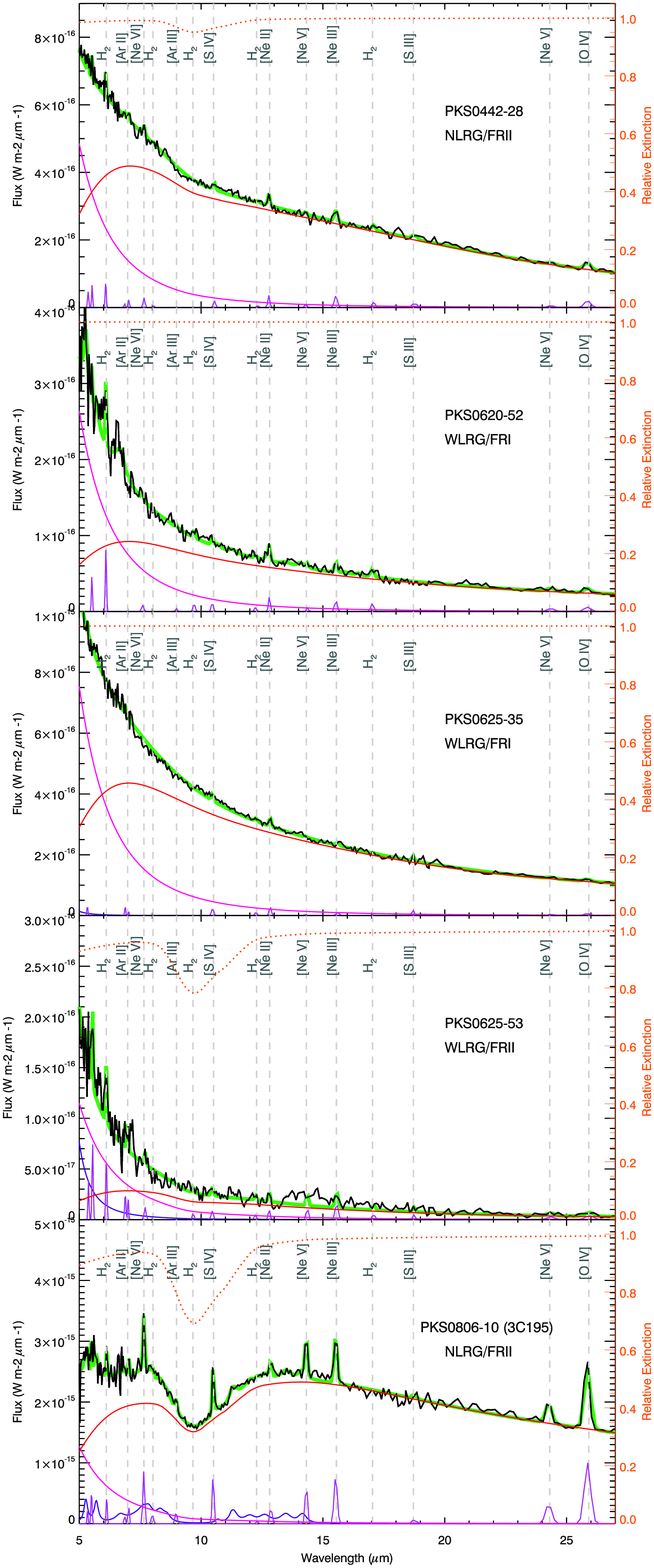}
\caption{Spitzer/IRS spectra for the 2Jy sample continued. Note that the data for PKS0404+03  were taken in Mapping Mode and were obtained from the Spitzer archive. \label{spec2} }
\end{figure*}

\begin{figure*}[t]
\epsscale{1}
\plottwo{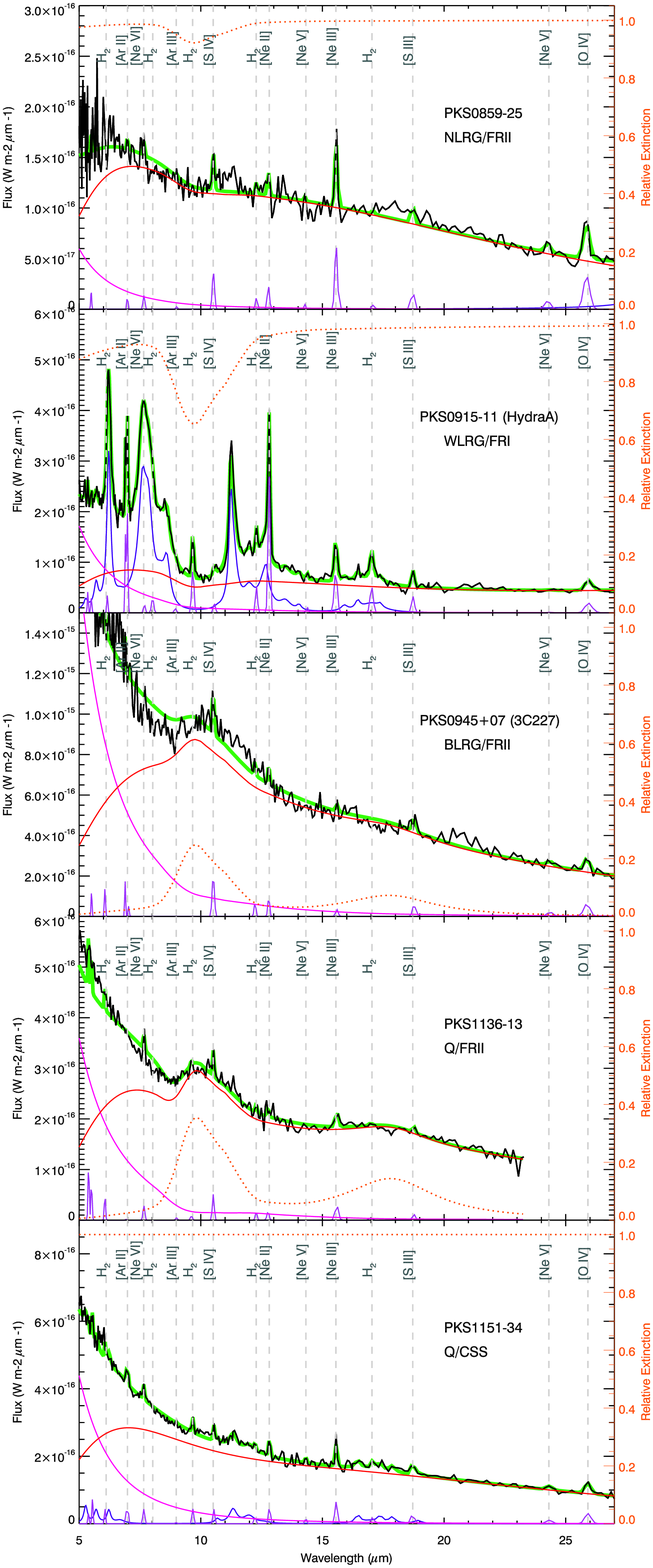}{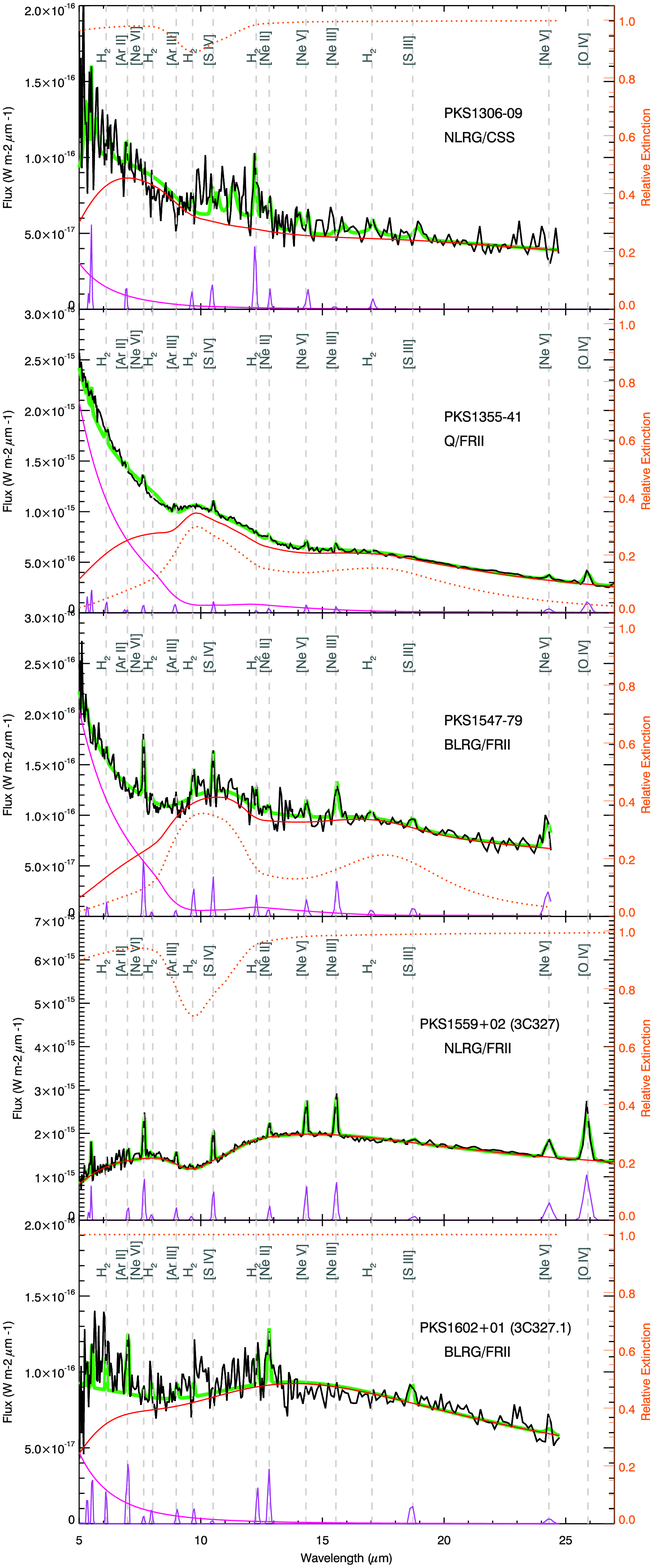}
\caption{Spitzer/IRS spectra for the 2Jy sample continued. Note that the data for PKS0915-11, PKS0945+07 and PKS1602+01 were obtained from the Spitzer archive, and the data for PKS0947+07 were taken in Mapping Mode. Potentially, the spectrum of PKS1602+01 suffers from enhanced flux calibration uncertainties at short wavelengths, due to saturation of the peak-up detector. {Hydra A also published in \citet{leipski09}.}  \label{spec3} }
\end{figure*}

\begin{figure*}[t]
\epsscale{1}
\plottwo{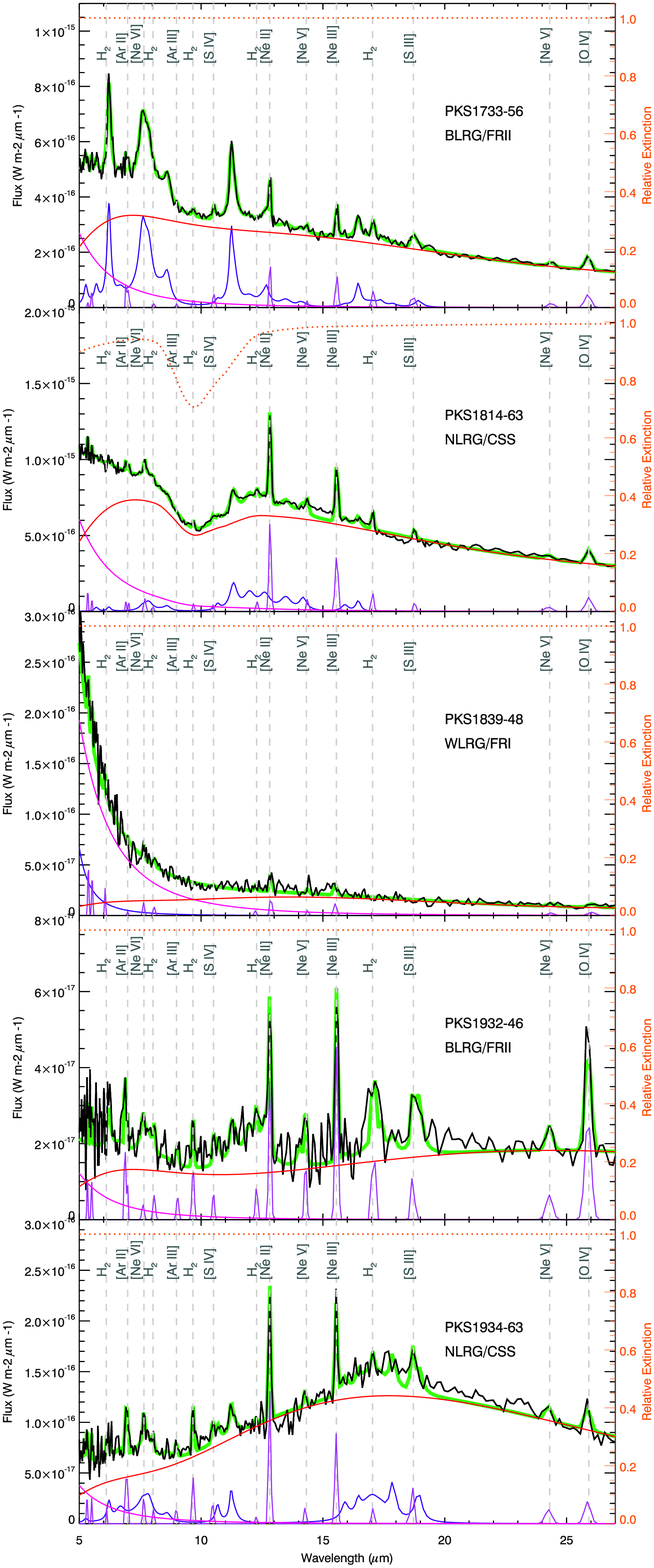}{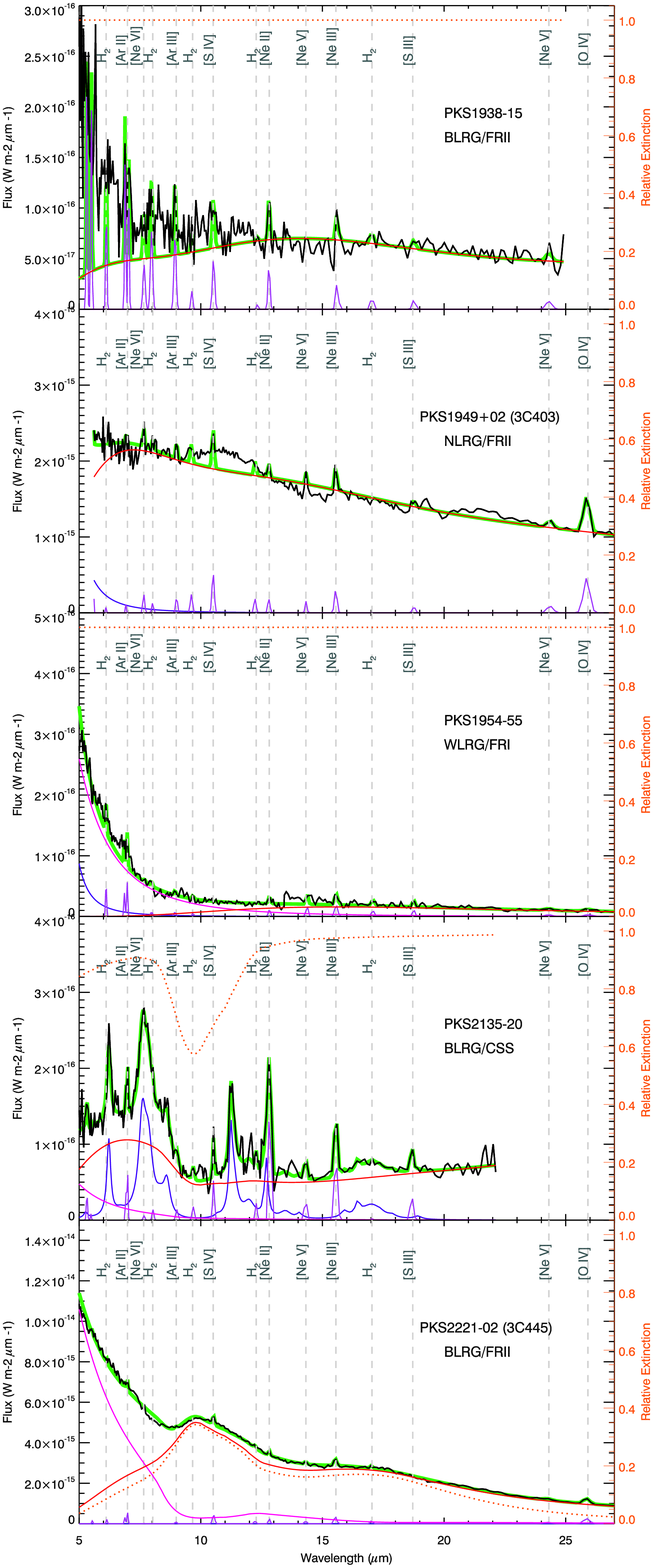}
\caption{Spitzer/IRS spectra for the 2Jy sample continued. Note that the data for PKS1949+02 (Mapping Mode) and PKS2221-02 were obtained from the Spitzer archive. \label{spec4} }
\end{figure*}

\begin{figure*}[t]
\epsscale{1}
\plottwo{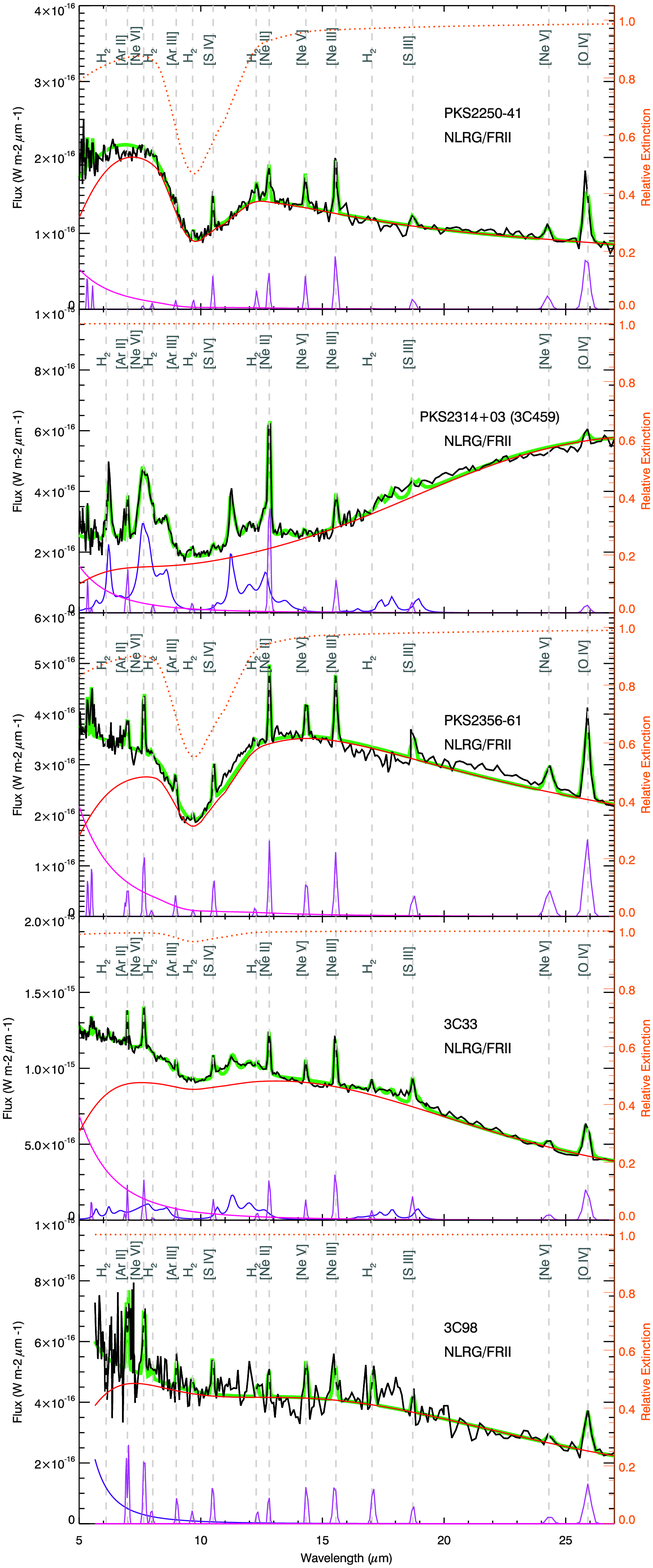}{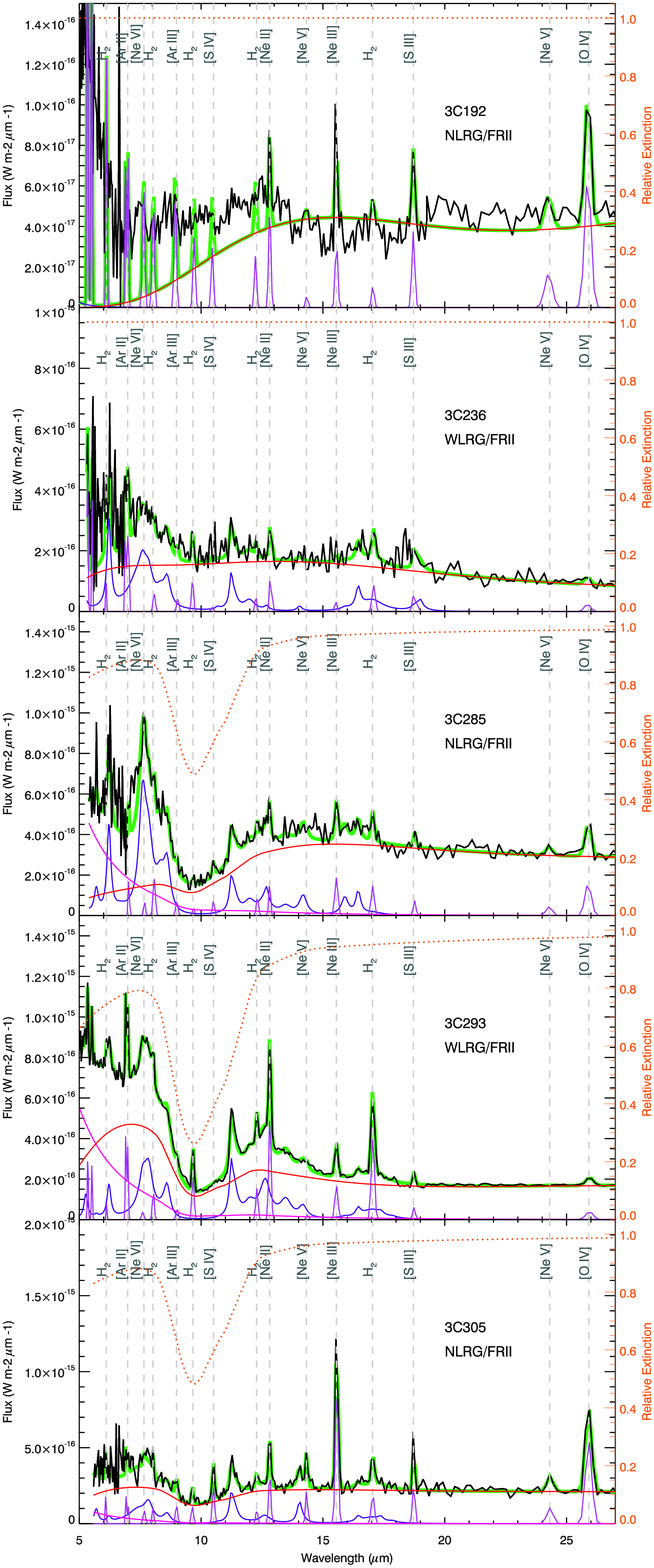}
\caption{Spitzer/IRS spectra for the 2Jy and 3CRR sample objects. All the 3CRR IRS data were obtained from the Spitzer archive. The data for 3C98, 3C236, 3C277.3, 3C285 and 3C305 were taken in Mapping Mode and were obtained from the Spitzer archive. {3C293 also published in \citet{leipski09} and 3C33 also published in \citet{ogle06}.} \label{spec5} }
\end{figure*}

\begin{figure*}
\epsscale{1}
\plottwo{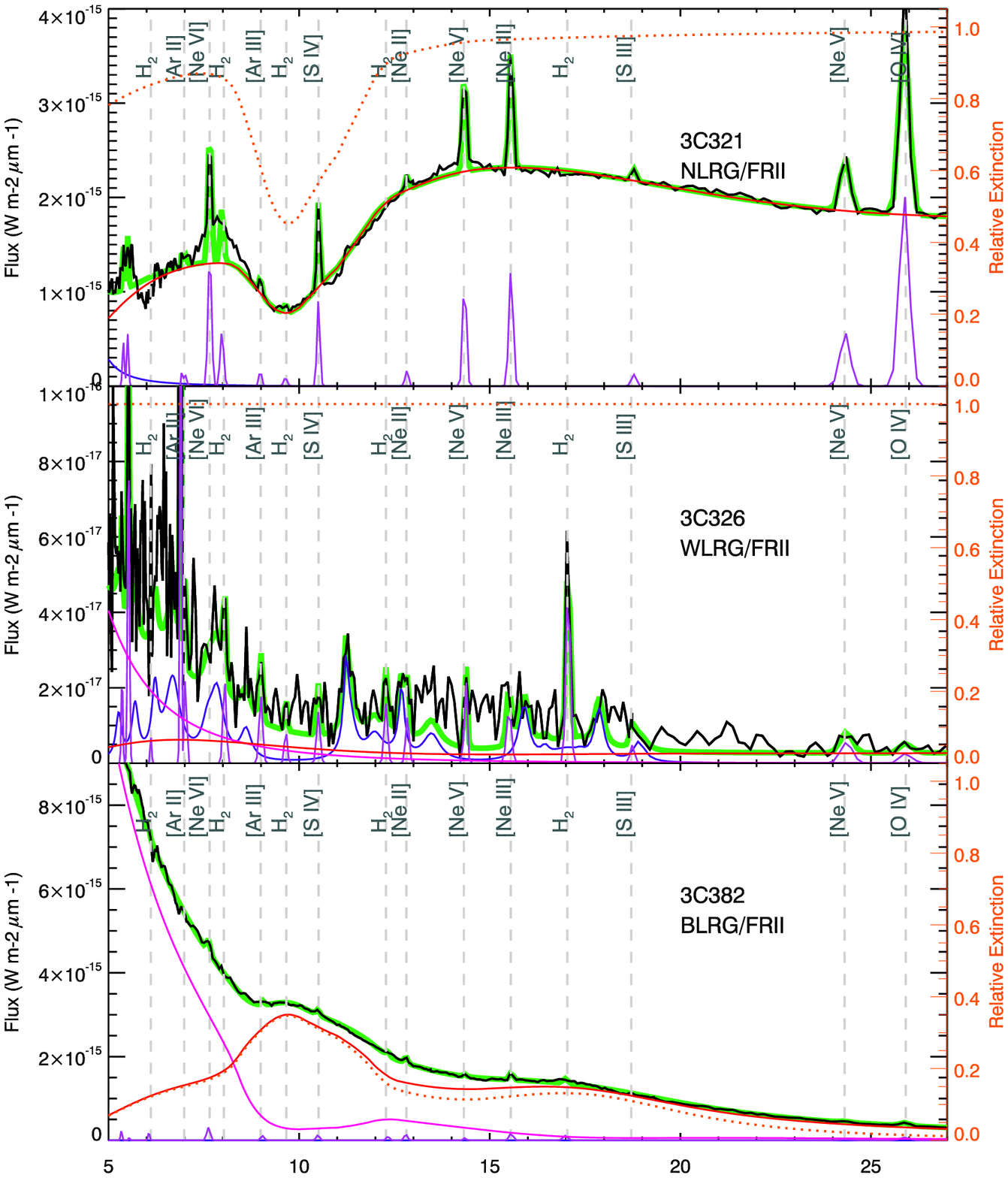}{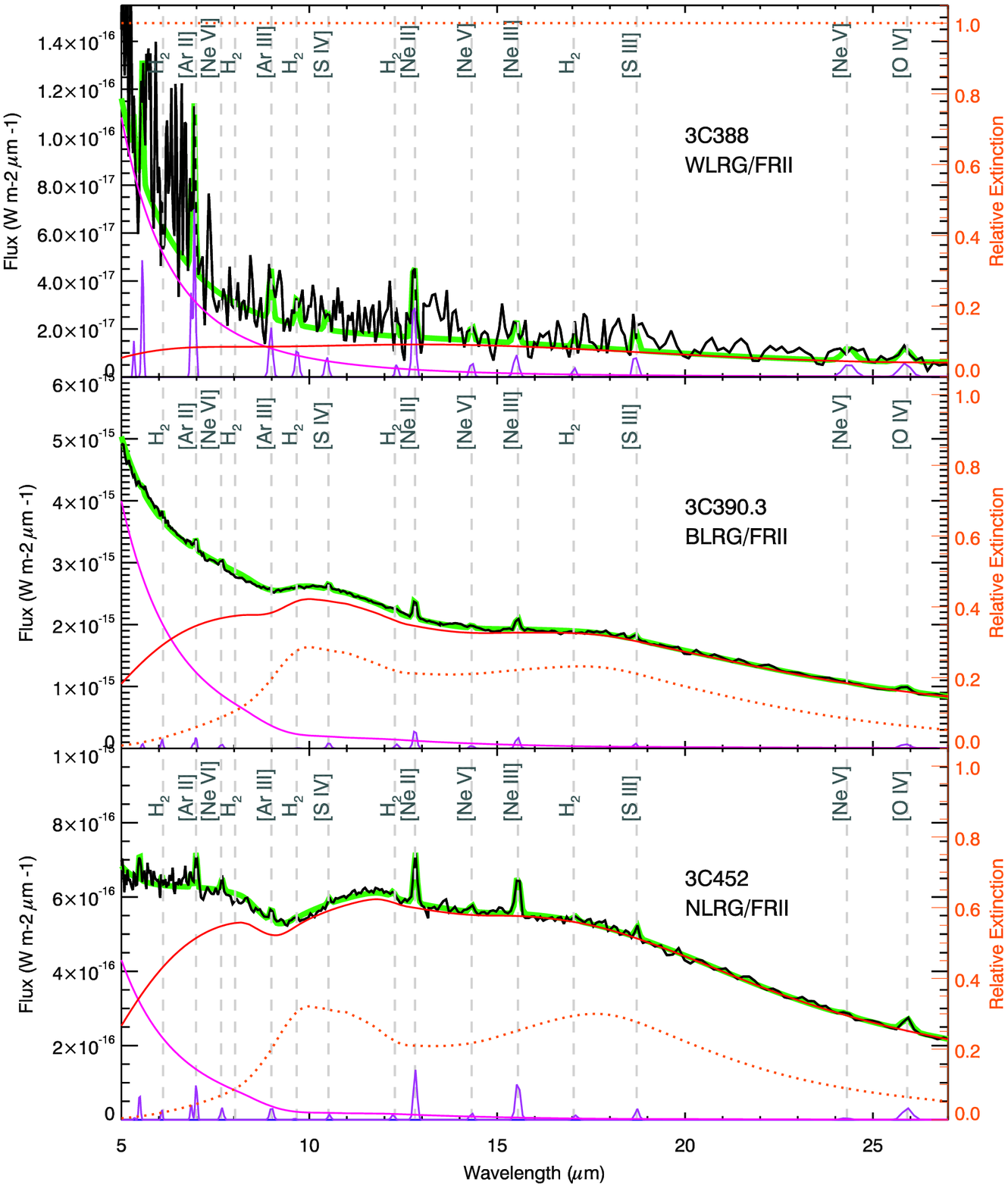}
\caption{Spitzer/IRS spectra of the 3CRR sample continued. \label{spec6} }
\end{figure*}

\end{document}